\documentclass[amssymb,11pt,toc,a4paper]{JHEP3}
\usepackage{cite}
\usepackage{epsfig}
\input epsf
\catcode `@=11 \@addtoreset{equation}{section} \catcode `@=12
\newcommand{\ba}{\begin{array}}
\newcommand{\ea}{\end{array}}
\newcommand{\be}{\begin{equation}}
\newcommand{\ee}{\end{equation}}
\newcommand{\bea}{\begin{eqnarray}}
\newcommand{\eea}{\end{eqnarray}}
\newcommand{\gsim}{\mathrel{\mathop{\kern 0pt \rlap
  {\raise.2ex\hbox{$>$}}} \lower.9ex\hbox{\kern-.190em $\sim$}}}
\newcommand{\lsim}{\mathrel{\mathop{\kern 0pt \rlap
  {\raise.2ex\hbox{$<$}}} \lower.9ex\hbox{\kern-.190em $\sim$}}}

\def\bbox{{\,\lower0.9pt\vbox{\hrule \hbox{\vrule height 0.2 cm
\hskip 0.2 cm \vrule height 0.2 cm}\hrule}\,}}
\newcommand{\dsl}{\pa \kern-0.5em /}

\newcommand{\nn}{\nonumber \\}

\def\* {&=&}
\def\d {{\rm d}}
\def\del {\partial}
\def\eq#1{(\ref{#1})}

\def\T{{\rm T}}
\def\U{{\rm U}}

\font\cmss = cmss12
\font\cmsss = cmss14

\def\integer{{\rlap{\cmss Z} \hskip 1.8pt \hbox{\cmss Z} }}
\def\identity{{\rlap{\rm I} \hskip 2.6pt \hbox{\rm I} }}
\def\tori{{\rlap{\cmss T} \hskip 2.0pt \hbox{\cmss T} }}
\def\laplace{{\kern1pt\vbox{\hrule height 1.2pt\hbox{\vrule width 1.2pt\hskip
  3pt\vbox{\vskip 6pt}\hskip 3pt\vrule width 0.6pt}\hrule height 0.6pt}
  \kern1pt}}
\def\roughly#1{\raise.3ex\hbox{$#1$\kern-.75em\lower1ex\hbox{$\sim$}}}

\def\real{{ \hbox{\cmss R} \llap{\vrule height 8pt width 0.8pt
depth -.1pt \hskip 0.6 pt \phantom. }}}

\def\tr{{\rm Tr}}
\def\S{{ \hbox{\cmss S} \llap{\vrule height 8pt width 0.6pt
depth -0.1pt \hskip -4.0pt \phantom. }}}
\def\H{\hbox{\cmss H}}
\def\U{\hbox{\cmsss U}}
\def\Z{\hbox{\cmsss Z}}

\title{\Large 
Quantum Aspects of GMS Solutions of Noncommutative Field Theory and
Large $N$ Limit of Matrix Models%
\thanks{Work supported in part by BK-21 Initiative in Physics 
(SNU - Project 2), KRF International Collaboration Grant, and KOSEF
Interdisciplinary Research Grant 98-07-02-07-01-5, KOSEF Leading
Scientist Program.}}
\author{Gautam Mandal ${}^a$, Soo-Jong Rey ${}^{b}$ 
            {\rm and} Spenta R. Wadia ${}^a$\\
$^a$ {\sl Department of Theoretical Physics,
 Tata Institute of Fundamental Research,\\
Homi Bhabha Road, Mumbai 400 005 \rm INDIA. }\\

$^b$ {\sl 
School of Physics \& Center for Theoretical Physics,\\ 
Seoul National University, Seoul 151-747 \rm KOREA}\\

email:\email{\\
mandal@theory.tifr.res.in
\\ sjrey@gravity.snu.ac.kr 
\\ wadia@theory.tifr.res.in}}

\preprint{\hepth{0111059}\\SNUST-011101\\TIFR-TH/01-32}

\abstract{We investigate quantum aspects of
Gopakumar-Minwalla-Strominger (GMS) solutions of noncommutative field
theory (NCFT) at large noncommutativity limit, $\theta \rightarrow
\infty$.  Building upon a quantitative map between operator formulation
of 2- (respectively, (2+1)-) dimensional NCFTs and large $N$ matrix
models of $c=0$ (respectively, $c=1$) noncritical strings, we show
that GMS solutions are quantum mechanically sensible only if we make
an appropriate joint scaling of $\theta$ and $N$.  For 't Hooft's
scaling, GMS solutions are replaced by large $N$ saddle-point solutions. 
GMS solutions are recovered from saddle-point solutions at small 't Hooft 
coupling regime, but are destabilized at large 'tHooft coupling regime by 
quantum effects. We make comparisons between these large $N$ effects and 
the recently studied infrared effects in NCFTs. We estimate the U(N) symmetry 
breaking effects of gradient term and argue that they are suppressed only 
at small 't Hooft coupling regime.}


\begin{document}

\section{Introduction}
Noncommutative field theories (NCFT), characterized by a noncommutativity
scale $\theta$, have been the subject of active research 
recently, largely because of their appearance in certain limits of the 
string theories and M-theory \cite{cds, sw}. 
These NCFTs deserve further study in its own 
right as they exhibit many properties which are elusive, if present at all,
in their commutative counterparts --- such as phenomenon of UV/IR mixing, 
T-duality 
and exact soliton/instanton (both BPS and non-BPS) solutions. One thus 
expects that a thorough understanding of NCFTs will shed new light on both
quantum field theories and string theories.

A step toward the understanding was provided by rich variety of classical
solutions. At large noncommutativity limit, $\theta \rightarrow \infty$, 
NCFT soliton/instanton solutions were constructed first by Gopakumar,
Minwalla and Strominger (GMS) \cite{gms}. Exact soliton/instanton solutions
were later constructed \cite{gms2} for finite noncommutativity, $\theta
< \infty$, as well. The classical
solutions have been studied in moduli space approximation
\cite{rocek, gopa2}, generalised to gauge theories \cite{ns, gn, poly, jatkar}, 
and applied to string theories in the context of tachyon condensation 
\cite{mukhi, harveyetal, wittentachyon, mandalrey}.

The emphasis of all these works were on finding the {\sl classical}
solutions, viz. the extrema of NCFT action. In this paper, we would like to 
address {\sl quantum-mechanical} solutions and their {\sl semiclassical limit},
equivalently, extrema of the functional integral (not just the action) 
of the NCFT. The first step to this goal would be to take into account the 
effect of the functional integral measure and study {\sl saddle-points}.
We then encounter a \underbar{puzzle} immediately. 

The simplest way to state the puzzle is as follow. Consider a NCFT
in Euclidean two dimensions, consisting of a scalar field $\T(x, y)$. 
In operator formulation, as defined by the Weyl-Moyal map, the field 
$\T(x,y)$ is represented by ${\bf T}$,  an ($\infty \times \infty$) matrix, 
equivalently, an operator in an auxiliary one-particle Hilbert space 
${\cal H}$. The formal similarity of the functional integral over 
${\bf T}$  to the matrix integral of a Hermitian $(N \times N)$ matrix  
\cite{bipz} is obvious. 

\begin{figure}[htb]
   \vspace{0.5cm}
\centerline{
   {\epsfxsize=5.5cm
   \epsfysize=5cm
   \epsffile{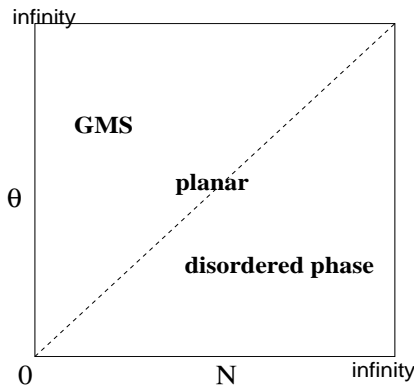}}
}
\caption{\sl Phases of two-dimensional noncommutative field theories. 
For $\theta \sim N^\nu$, the GMS-, planar-, and disordered phases correspond to 
$\nu > 1, = 1, < 1$, respectively.}
\label{wilson}
\end{figure}
An important point to note is that, in the one-matrix model,
the measure of matrix integration, the famous `Coulomb 
repulsion' term, 
changes the classical vacuum dramatically \cite{matrixmodel}. 
Indeed, the measure effect, which scales as ${\cal O}(N^2)$, dominates over the 
classical action, which 
scales naively as ${\cal O}(N)$ {\sl unless} a suitable scaling of 
coupling parameters in the classical action is made. As we will show in 
Section 2, in the two-dimensional Euclidean NCFTs, 
the only way the classical 
action can compete with the measure effect is to take a large-$\theta$ limit 
in an appropriate way. Specifically, in the large-$\theta$ limit, the 
quantum effective action is given schematically as:
\bea 
S_{\rm eff}[\theta, N] = S_{\rm classical}[\theta, N] + S_{\rm measure}[N],
\label{theaction}
\eea 
where
\bea
S_{\rm classical}[\theta, N]  \quad \sim \quad {\cal O}(\theta N)
\quad {\rm and} \qquad
S_{\rm measure} \quad \sim \quad {\cal O} (N^2) .
\nonumber
\label{class-vs-measure} 
\eea 
Clearly, there are different ways of taking a large $\theta$,
large-$N$ limit, leading to three distinct phases: 
\bea
{\rm (a)} \,\,\,\,\,\, \quad {\rm GMS~~phase} \,\,: \qquad
\theta \, &\sim& \, N^{\nu} \to \infty
\qquad (\nu > 1) \nonumber \\
{\rm (b)} \,\,\,\quad {\rm planar~~phase} 
\,\, : \qquad \theta \, &\sim& \, 
N \to \infty \qquad \quad g^2_{\rm eff} = {\rm fixed} \nonumber \\
{\rm (c)} \,\,\,\,{\rm disordered~~phase} : \qquad
{\rm }\theta \, &\sim& \, N^{\nu} 
\to \infty \qquad (\nu < 1).
\label{3limits}
\eea
Evidently, it is only with the scalings (a) and (b) the classical
action can compete with the term coming from the measure effect. In
the limit (a) the classical term dominates, therefore the GMS
solutions remains a good quantum solution. The case (b) turns out 
to be equivalent to the `t
Hooft planar limit (see Sec 2); in this case the measure term and the
classical action are comparable, implying that the {\sl saddle point
solutions are different from the GMS solutions}.  In case (c), or for
a fixed $\theta$ as is assumed for {\sl classical} NCFT instantons,
the measure effect $S_{\rm measure}$ becomes infinitely larger than
the classical action $S_{\rm classical}$ and indeed seems to drive
system to a different phase, referred as disordered phase, altogether.

The aforementioned three phases exist also for {\sl quantum} vacua and
solitons in $(2+1)$-dimensional NCFTs, although the way the functional
integral measure effects come about is somewhat different. Evaluating
the energy for vacua and solitons, we argue that quantum corrections
are small for GMS-phase, but become sizable for planar- and
disordered phases. In particular, in disordered phase, we find an 
indication that the classical vacua and solitons are destabilized 
completely once the measure effects are taken into account.

The paper is organised as follows. In section 2, we analyze the
above results for two-dimensional Euclidean NCFT as an appropriate limit 
of the Hermitian one-matrix model \cite{bipz} studied previously in the
context of $c<1$ noncritical strings \cite{c<1}. 
In section 3, we provide both perturbative and nonperturbative 
estimates of the gradient effect, which were dropped in the analysis of
section 2. In section 4, we extend the consideration 
to $(2+1)$ dimensional NCFT by studying its matrix model
analog, viz. the time-dependent Hermitian-matrix model studied 
previously in the context of $c=1$ noncritical string \cite{c=1}. 
Among the interesting consequences caused by quantum fluctuations, 
we point out spontaneous breakdown of translation invariance, and 
decrease of the soliton mass. In the last section, we remark briefly 
concerning possible relevance of the results to IKKT \cite{ikkt} and 
BFSS \cite{bfss} matrix models, and to the phenomenon of the UV-IR mixing
\cite{uvir}.

A preliminary version of this work was presented in \cite{Rey-talk}.

\section{Two-Dimensional Noncommutative Field Theories}
\subsection{Classical Theory}
Begin with noncommutative plane $\real^2_\theta$, whose coordinates
${\bf y}$ obey the Heisenberg algebra:
\bea
\Big[ y^a, y^b] = i \theta^{ab} = i \theta \epsilon^{ab}, 
\qquad (a,b = 1, 2). 
\label{nc2algebra}
\eea
We shall be studying a Euclidean field theory on $\real^2_\theta$, 
consisting of a scalar field $\T({\bf y})$ with self-interaction potential
-- in general polynomial -- $V(\T)$. 
Via the Seiberg-Witten map \cite{sw}, 
the theory is describable equivalently in terms 
of a noncommutative field theory (NCFT) on $\real^2$, whose action is 
given by:
\begin{eqnarray}
S_{{\rm NC}}[\theta; V] = \int_{\real^2} \! \d^2{\bf y} \, 
\left[ {\frac{1 }{2}} \partial_{{\bf y}} \T \star_\theta \partial_{{\bf y}} \T 
+ V_{\star_\theta} (\T) \right].  
\label{2daction}
\end{eqnarray}
In NCFT, the noncommutativity $\theta^{ab}$ is encoded through the 
$\star_\theta$-product: 
\begin{eqnarray}
\Big(A \star_\theta B\Big)({\bf y}) := 
\exp \left( {i\over2} \theta^{ab} \partial_{{\bf y}_1}^a
\wedge \partial_{{\bf y}_2}^b \right) A ({\bf y}_1) \, B({\bf y}_2)
|_{{\bf y}_2= {\bf y}_1= {\bf y}}
\label{moyalprod}
\end{eqnarray}
 
It has been noted that a theory of the type Eq.(\ref{2daction}) arises for the
level-zero truncation of the open string field theory on Euclidean worldvolume
of an unstable D1-brane, either in bosonic or in Type IIA string theories, 
on which a nonzero, constant background of the (Euclideanized) two-form 
potential $B_2$ is turned on \cite{sentachyon}. The scalar field 
$\T({\bf x})$ in Eq.(\ref{2daction}) represents, when expanded around top 
of the potential $V(\T)$, the real-valued tachyon field in these situations.

Inverse of the noncommutativity parameter, $1 / {\rm }\theta$,
plays the role of a coupling paramter of the NCFT. To see this, rescale the
coordinates as:
\bea
{\bf y}\quad \rightarrow \quad {\bf x}\,\,=\,\,{1 \over \sqrt{\theta} }
{\bf y} \qquad {\rm so} \quad {\rm that} \qquad
\left[ x^a, x^b \right] = i \epsilon^{ab}
\nonumber
\eea 
and expand the NCFT action Eq.(\ref{2daction}) in powers of 
$(1/{\rm }\theta)$:
\bea
S_{{\rm NC}}[\theta; V_\star]
={\rm } \theta \int_{\real^2} \d^2{\bf x}\,\left[ {\cal L}%
_0+{1 \over {\rm } \theta } {\cal L}_{-1}+\cdots \right].
\label{expand2daction}
\eea
Here,
\bea
{\cal L}_0=V_{\star}(\T)\qquad \quad {\rm and}\quad \qquad {\cal L%
}_{-1}={\frac 12}\left( \partial _{{\bf x}} \T\right) ^2, 
\label{twoterms}
\eea
and the $\star$'s refer to the Moyal-product Eq.(\ref{moyalprod}) in which the
noncommutativity parameter 
$\theta^{ab} $ is replaced by $\epsilon^{ab}$. Evidently, at large
noncommutativity, $(1/{\rm }\theta) \rightarrow 0$, the gradient-term 
${\cal L}_{-1}$ yields a sub-leading order correction \footnote{Quantum 
mechanically, somewhat surprisingly, the ${\cal L}_{-1}$ term contributes 
leading-order effects in the planar expansion in powers of $1/N$. 
In section 3, we will show that small `t Hooft coupling suppresses the 
contribution compared to those from the ${\cal L}_0$ term}.

Utilizing the Weyl-Moyal map (See Appendix A), one can map the two-dimensional 
NCFT Eq.(\ref{2daction}) to a {\sl zero}-dimensional Hermitian matrix model,
defined by
\bea
\S_{\rm NC} [\theta; V] ={\rm } \theta \,{\tr}_{{\cal H}}
\left[ \,V({\bf T}) + {1 \over {\rm }\theta} \left( -{1 \over 2}
\left[ \hat{\bf x}, {\bf T} \right]^2 \right) +\cdots
\,\right]. 
\label{weylaction}
\eea

\subsection{Classical Vacua and Instantons}
Classical solutions of the NCFT are most straightforwardly obtainable from
Eq.(\ref{weylaction}). At leading order in 
${\rm } \theta \rightarrow \infty$, the classical solutions are critical 
points of the potential,  $V' ({\bf T}) = 0$, viz. a matrix-valued 
algebraic equation of degree-$(P-1)$. Denote local minima of the polynomial 
function $V(\lambda)$ as $\lambda_0, \lambda_1, \lambda_2, \cdots$, 
conveniently labelled in ascending order: $V(\lambda_0) \le V(\lambda_1) \le 
V(\lambda_2) \le \cdots$.

One then finds that the most general classical solution of 
$V^{\prime}({\bf T}) = 0$ takes the form: 
\begin{eqnarray}
{\bf T} = \sum_{\ell =1}^N \lambda_{a_\ell} {\bf P}_\ell  \nonumber
\label{gen-sol}
\end{eqnarray}
where $\lambda_{a_\ell}$'s take values out of the set $(\lambda_0,
\lambda_1, \cdots)$ permitting duplications. 
We will define eigenvalue density $\rho(\lambda)$ as
\bea
\rho(\lambda) := {1 \over {\tt dim}{\cal H}}
\sum_a \delta(\lambda - \lambda_a).
\label{density}
\eea
As a concrete example, consider a symmetric double-well
potential:
\bea 
V({\bf T}) = V_0 + \frac{\lambda_4}{4} \left( {\bf T}^2 - \T_0^2 \right)^2,
\label{doublewell}
\eea
for which the roots $\lambda_a$ are $\pm \T_0$. 

{\bf Vacua:}

Using Eq.\eq{gen-sol}, we easily find the doubly degenerate vacua (R and
L, for left and right), given by
\bea
{\bf T}_{\rm \small R,L} = \pm \T_0 \identity,
\label{vacua}
\eea
These solutions are exact and are valid for any $\theta$, small 
or large. The energy $E_0$ is given by
$E_0 = (N\theta)V_0$.

{\bf Instantons:}

The other solutions, using Eq.\eq{gen-sol}, are given by 
\begin{eqnarray}
{\bf T}_{N_1, N_2} = \T_0 \left( {\bf P}_{[N_1]} - {\bf P}_{[N_2]} \right).
\label{instantons}
\end{eqnarray}
These solutions are generally valid only at large $\theta$, with 
${\cal O}(1/\theta)$ corrections affecting both their profile as well as 
their energy. The notation ${\bf P}_{[N_1]}$ stands for a 
projection operator of rank $N_1$, similarly for $N_2$. 
We will call the solution Eq.\eq{instantons} an 
`$(N_1, N_2)$-instanton'.

>From Eq.\eq{density}, we find that the above vacua and instantons 
yield the following density profiles: 
\begin{eqnarray}
\rho_{\rm \small ~R,~L} \, (\lambda) &=& \delta(\lambda \mp \T_0)
\nonumber \\
\rho_{[N_1, N_2]} (\lambda) &=& n_1 \delta (\lambda - \T_0) + n_2 \, 
\delta(\lambda + \T_0) \qquad {\rm where} \qquad  n_{1,2} 
= {\frac{N_{1,2} }{{\tt dim}{\cal H}}}.  
\label{densitysolution}
\end{eqnarray}

\subsection{Quantum Theory}
\subsubsection*{Definition}
The quantum NCFT is defined via the following \underline{regularized} 
partition function:
\bea
{\cal Z}_{{\rm NC}}[\theta, V_\star;L_1 L_2] \,= \,
\int \left[ \d \T \right]_{L_1, L_2} 
\,\exp \Big( - S_{{\rm NC}}[\theta;V_\star] \Big),
\label{moyalpartition} 
\eea
where $S_{{\rm NC}}$ is given by Eq.\eq{expand2daction}.
Here $L_1, L_2$ represent large distance cut-offs 
introduced as regulator of possible infrared divergences. Generically, the
theory also needs an ultraviolet cut-off, e.g.
a lattice spacing $a$; the theories discussed in this paper
will be taken ultraviolet-renormalizable. We will assume that, in
the above definition Eq.(\ref{moyalpartition}), 
the limit $a\to 0$ has been taken. 

In the previous section, we have seen that classical NCFT is equivalent,
via the Weyl-Moyal correspondence, to a model of a $(\infty \times \infty)$ 
Hermitian matrix, Eq.(\ref{weylaction}). What then would be the 
corresponding statement at quantum level? As 
the theory Eq.\eq{moyalpartition} is {\sl defined} 
with the large distance
cutoff's $L_1, L_2$, one is naturally led to a noncommutative
torus (see, e.g.\cite{cds}) as a concrete setup for infrared regularization.
See Fig.\eq{fig2} for an illustration.
\begin{figure}[htb]
   \vspace{0.5cm}
\centerline{
   {\epsfxsize=11.5cm
    \epsfysize=5cm
    \epsffile{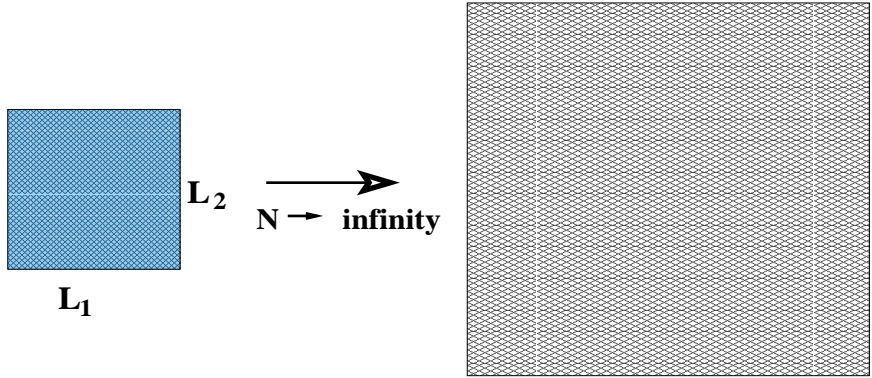}}
}
\caption{\sl Noncommutative plane as a continuum and 
large-volume limit of noncommutative torus. The limit requires $N \rightarrow
\infty$.}
\label{fig2}
\end{figure}

Start with a noncommutative torus $\tori^2_\theta$, defined through so-called
quotient condition on $(N \times N)$ matrices $X^1, X^2$ as:
\bea
X^a + L^a \delta^a_b \identity_N =\U_b^{-1} X^a \U_b \qquad \quad
(a, b = 1, 2).
\nonumber 
\eea
Generically, a nontrivial solution to the quotient condition requires 
$N \rightarrow \infty$. Applying the condition on two different directions 
on 
$\tori^2_\theta$, one finds that the quotients $\U_a$'s ought to obey
\bea
\U_a  \U_b \, \U^{-1}_a \U^{-1}_b = e^{ - i \Theta^{ab}} \, \identity.
\nonumber
\eea
where $\Theta_{ab}$ is dimensionless and will
shortly be identified, for a square torus,  
with $\Theta^{ab}= \theta^{ab} (\frac{2\pi}L)^2, \ (L_1= L_2 = L)$.

The scalar field $\bf T$ defined on $\tori^2_\theta$ is defined via
\bea
{\bf T} = \sum_{\{{\bf m}\}= \integer} \widetilde{\bf T}_{\bf m} \, 
\U_1^{m_1} \, \U_2^{m_2}.
\nonumber
\eea
Here, $\widetilde{\bf T}_{\bf m}$ belongs to a sufficiently rapidly 
decreasing sequence of appropriate Schwartz space. 

One can represent the  $\U_1, \U_2$ basis in terms of Hermitian 
operators of
the form:
\bea
\U_a = \exp \left( 2 \pi i {{\bf y}^a \over L^a} \right) \qquad \qquad 
(a = 1, 2),
\nonumber
\eea
where, in the large-$N$ limit,
\bea
\left[\widehat{y^a}, \widehat{y^b} \right]  
\approx i \left( {L_a L_b \over (2 \pi)^2} \right) \Theta^{ab} 
\equiv i\ \theta^{ab}
,
\nonumber
\eea
in which we have used Eq.\eq{nc2algebra} in the last step.

The simplest situation arises for 
so-called rational noncommutative tori. For
our purposes it is sufficiently general to consider,
among these, the case when 
\be \Theta^{ab} = \Theta \epsilon^{ab}
\qquad {\rm and} \qquad
\ \Theta= 1/N
.
\nonumber
\ee
Focusing on the square torus $L_1= L_2 = L$ from now on,
we get, using the above two equations,
\be
\theta = \left( \frac{L}{2\pi}
\right)^2 \Theta = \left(\frac{L}{2\pi}\right)^2 
\frac1{N}
.
\label{map}
\ee
For the noncommutative torus with such a value of $\theta$ , 
the Weyl-Moyal correspondence maps the partition function
Eq.(\ref{moyalpartition}) of the NCFT on $\tori^2_\theta$ to 
the following partition function for a Hermitian matrix of size 
$(N \times N)$:
\bea
{\Z}_{N}[\Theta, V; N] \, = \,
\int \left[ \d {\bf T} \right]_N
\, \exp \Big( - \S_{{\rm NC}}[\Theta; V({\bf T})] \Big),
\label{weylpartition}
\eea
where the matrix integral measure is given by
\bea
\left[ \d {\bf T} \right]_N := \prod_{i = 1}^N
d {\bf T}_{ii} \prod_{1 \le i < j \le N} 2 
\d \, {\rm Re} ({\bf T}_{ij}) \d \, {\rm Im} ({\bf T}_{ij}).
\nonumber
\eea
Let us 
now consider the limit $L \rightarrow \infty$;
in this limit the noncommutative torus 
$\tori^2_\theta$ ought to approach the noncommutative plane $\real^2_\theta$.
Since the Heisenberg algebra Eq.\eq{nc2algebra} on
$\real^2_\theta$ has only
infinite-dimensional representations, the above limit must
also be accompanied by a limit $N \rightarrow \infty$.
As $\theta \sim L^2/N$ (from Eq.\eq{map}), 
the large-$\theta$ limit discussed in Sec 2.1 can be
attained by
\bea
L \to \infty, \qquad N \to \infty, \qquad {\rm and} \qquad
\theta \sim L^2/N \to \infty.
\nonumber
\eea
This is achievable by letting 
\bea
L \sim N^\gamma \quad \Rightarrow \quad \theta \sim N^\nu
\qquad {\rm where} \qquad \nu = (2\gamma-1),
\label{scaling}
\eea
where we assume
\bea
\gamma> \frac12 \quad \Rightarrow  \quad \nu > 0. 
\nonumber
\eea
To sum up, the above observations lead to the definition of the quantum NCFT 
on $\real^2_\theta$ as follows:
\bea
\nonumber \\
\lim_{L_1 L_2 \rightarrow \infty} \, 
{\cal Z}_{\rm NC} [\theta, V_\star; L_1 L_2] 
\quad
\equiv 
\quad
\lim_{N \rightarrow \infty} \,
\Z_N [\theta, V; N],
\nonumber \\
\label{definition}
\eea
where, on the right-hand side, the noncommutativity parameter $\theta$
is given in terms of Eq.(\ref{map}).

\subsection{Classical, Planar, and Disordered Phases of NCFT$_2$} 
The Weyl-Moyal equivalence Eq.(\ref{definition}), together
with Eq.\eq{scaling},  indicates that the quantum 
NCFT is actually defined in terms of a double-series expansion: 
large-$N$, and
large-$\theta$ expansions. 
To detail, define the quantum NCFT in terms of the 
Hermitian matrix model, as in the right-hand side of Eq.(\ref{definition}).
Suppose, at large $\theta$,  we ignore the subleading part ${\cal L}_{-1}$ 
in the action Eq.\eq{expand2daction}. In that case, the partition function 
becomes identical in form to
the one-matrix integral \cite{bipz} and the $c<1$ matrix models for
$c<1$ noncritical strings \cite{c<1}. These latter models are defined 
in terms of the matrix model partition function $\integer_{\rm mm}$ 
\be
\integer_{\rm mm}[\beta,V; N]= \int \d {\bf M} \, 
\exp \Big(-\beta {\rm Tr}_N V({\bf M}) \Big)
,
\label{matrixpartition}
\ee
where $V(x)$ is the Boltzmann function, taking a polynomial form:
$V(x) = a_2 x^2 + a_4 x^4 + \ldots$. Evidently, modulo the identification
,
$\theta = \beta$, 
we have
\be
\Z_N [\theta, V; N] \quad = \quad \integer_{\rm mm}[\beta,V; N].
\label{analogy}
\ee
To investigate the partition function $\Z_N$, we will
therefore proceed as in the case of the one-matrix model.
Integrating out the `angular part' of ${\bf T}$, the partition function 
$\Z_N$ is rewritable as an integral over the $N$ eigenvalues 
$\lambda_1, \lambda_2, \cdots, \lambda_N$ of ${\bf T}$ \cite{bipz}:
\bea
\Z_N [\theta, V; N]
&=& C_N \int \prod_{k=1}^N \d \lambda_k 
\prod_{k < \ell}^N (\lambda_k - \lambda_\ell)^2 \,
\exp \Big( - {\rm }\theta \sum_k V(\lambda_k) \Big)
\nonumber \\
&=& C_N \int \prod_{k=1}^N \d \lambda_k
\exp \Big( -  S_{\rm eff} 
(\lambda_1, \cdots, \lambda_N) \Big),
\label{partitiondef}
\eea
where
\bea
C_N = {\rm Vol} \left[ { {\rm U}(N) \over {{\rm U}(1)}^N \times
{\cal S}_N} \right] = {1 \over N!} 
\prod_{K=1}^N {(2 \pi)^{K-1} \over \Gamma(K)}
\nonumber
\eea
refers to the angular volume measure factor, and
\bea
S_{\rm eff} (\lambda_1, \cdots, \lambda_N)  \,\,
= \,\, S_{\rm classical} [N, \theta]
\, + \, S_{\rm measure}[N] 
\label{effaction}
\eea
with
\bea
S_{\rm classical}[N, \theta]
= N {\rm } \theta \left( {1 \over N} \!\! \sum_{k=1}^N V(\lambda_k) \right)
 \qquad
{\rm and}
\qquad
S_{\rm measure} [N] =  N^2 \left( 
{1 \over 2 N^2} \!\!\! \sum_{1 \le k \ne \ell \le N} 
\ln (\lambda_k - \lambda_\ell)^2 \right)
\nonumber 
\eea
refers to the effective action as a sum of the classical contribution
and the measure factor contribution. 

The large-$N$ limit of one-matrix
models are describable by a {\sl master field}
configuration, where distribution of 
the eigenvalues $\lambda_1, \cdots, \lambda_N$ are
encoded into the density field $\rho(\lambda)$, introduced in 
Eq.(\ref{density}), with support over connected 
compact domains ${\cal D}$ and subject to the constraints:
\bea
\int_{\cal D} \d \lambda \rho(\lambda) = 1 \qquad
\quad {\rm and} \quad \qquad \rho(\lambda) \ge 0 \quad
{\rm on} \quad \lambda \subset {\cal D}.
\label{constraint}
\eea
The effective action of eigenvalues then become
\bea
S_{\rm eff} [\rho]	
= N^2 \left[
g^{-2}_{\rm eff}
\int_{\cal D} \!\d \lambda \, \rho(\lambda) V(\lambda) 
- \int_{\cal D} \! \d \lambda 
\!\int_{{\cal D}} \! \d \mu \, 
\rho(\lambda) \Big( \ln \vert \lambda - \mu \vert \Big) \rho(\mu) \right],
\label{collaction}
\eea
in which
\bea
g^2_{\rm eff} \equiv {N \over {\rm } \theta}
\label{thooftcoupling}
\eea
measures the relative weight between the classical contribution and
the measure factor contribution. 

Now the effective action Eq.(\ref{collaction}) is exactly of the form 
as Eq.\eq{theaction}.
One thus discovers that, in {\sl quantum} NCFT, there ought to
exist three distinct regimes as in Eq.\eq{3limits}.
If one were to define the quantum NCFT in terms of the Hermitian 
matrix model, as in Eq.(\ref{definition}), via Weyl-Moyal equivalence,
the three different regimes are distinguished by relative 
weight in Eq.\eq{collaction} 
between the classical contribution $S_{\rm classical} 
\sim {\cal O}(N {\rm }\theta)$ and the matrix-integral measure part
contribution $S_{\rm measure} \sim {\cal O}(N^2)$.

The above considerations entail an important consequence to the
interpretation of the noncommutative field theories and the classical 
solutions therein, as studied in \cite{gms}. 
First, in noncommutative field theory, one 
defines the theory by viewing noncommutative field ${\T}$ as a 
representation of the Heisenberg algebra, which is infinite-dimensional
in case the theory is defined on $\real^2_\theta$. If one interprets this
as meaning that the size $N$ of the matrix-field ${\bf T}$ is strictly
infinite to begin with, then the classical action $S_{\rm classical}$ 
becomes insignificant, as it is far outweighted by the quantum contribution
$S_{\rm measure}$ coming from the matrix-interal measure. Second, in order 
to be able to view the classical solutions, e.g. solutions studied in 
\cite{gms}, as saddle-points of the partition function 
Eq.(\ref{moyalpartition}), one must first `regulate' the noncommutative 
field theory in such a way that the corresponding Weyl formulation
is defined on a finite $N$-dimensional Hilbert space to begin with, viz.
the Hermitian matrix model is for $(N \times N)$ matrices. 
In order to recover a sensible saddle-point solution, one {\sl subsequently} 
needs to take an appropriate large-$\theta$, large-$N$ limit. Eq.(\ref{3limits}) indicates that, a priori, there are three types of possible scaling 
of the noncommutative field theory.
Based on this observation, we thus conclude that, only in the classical
scaling (a), the classical solutions found in \cite{gms} are also the 
saddle-point solutions. For the planar scaling (b), the classical solutions 
ought to be replaced, as we will find in the next section, by new ones in which 
the eigenvalues are distributed. In the quantum scaling (c), the classical
solutions found in \cite{gms} are washed out completely. 

\subsection{Quantum Vacua and Instantons}
We now flesh up the preceding discussion by studying the
{\sl quantum} vacua and instantons of the two-dimensional NCFT on
$\real^2_\theta$ in the large-$N$, large-$\theta$ limit in the various
regimes Eq.\eq{3limits}. In doing so, we will use the analogy Eq.\eq{analogy} 
with the one-matrix model studied in the context of $c<1$ noncritical string
\cite{c<1} to quantize the solutions described in Sec 2.2. We will do explicit
calculations in the GMS- and planar-phases and will make some
qualitative remarks about the disordered phase.

We begin by defining the `free'energy $F[\theta, V; N]$  as:
\bea
\Z_N[\theta, V; N] := \left({ 2 \pi \over N} \right)^{N^2/2} 
\, e^{-F[\theta, V; N]},
\nonumber 
\eea
where 
the normalization is chosen so that $F = 0$ for quadratic potential
$V({\bf T}) = {1 \over 2} {\rm Tr} {\bf T}^2$.
As is well-known \cite{bipz}, the free energy  
has the following large-$N$ expansion:
\be
F[\theta= (N/g_{\rm eff}^2), N]= N^2 F_0(g^2_{\rm eff}) + F_1(g^2_{\rm eff})
+ N^{-2} F_2(g^2_{\rm eff}) + \ldots,
\label{bipz-expansion}
\ee
where each of the $F_n$ are defined via power series in $g^2_{\rm eff}$ with
a radius of convergence $g_c$. It will be convenient at
this stage to rephrase the three limits Eq.\eq{3limits}  as 
\bea 
\begin{array}{cccc}
{\rm (a)} \hskip0.5cm &  {\rm GMS~~phase} : & \hskip0.5cm N\to \infty, &\ 
\hskip0.5cm g_{\rm eff} \to 0
\\
{\rm (b)} \hskip0.5cm & {\rm planar~~phase} : & \hskip0.5cm N\to \infty, & \ 
\hskip0.92cm g_{\rm eff} = {\rm fixed}
\\
{\rm (c)} \hskip0.5cm & {\rm disordered~~phase} : 
& \hskip0.5cm N\to \infty, & \ 
\hskip0.65cm g_{\rm eff} \to \infty
.
\end{array}
\label{3limits1}
\eea
The leading term $F_0$ in Eq.\eq{bipz-expansion} is given 
by the saddle-point contribution at large-$N$ limit. Clearly,
the leading behavior of (a) the GMS-phase free energy, and 
(b) the planar-phase free energy (for $g_{\rm eff}< g_c$) 
are derivable from this saddle-point expression. 
The disordered phase free energy is clearly in the strong coupling
phase $g_{\rm eff} > g_c$ in which the large-$N$ expansion 
Eq.\eq{bipz-expansion} breaks down.   

We see, therefore, that we can derive the leading behaviour of the
partition function $\Z_N$ in the double limit, 
$N\to \infty, \theta\to \infty$, from the large-$N$ saddle-point (except in the disordered phase).
We describe in Appendix B how to compute the large-$N$ saddle-point 
as minima of the effective action
Eq.\eq{collaction} subject to the constraint Eq.\eq{constraint}.
We simply quote the result here (see Appendix B or
\cite{bipz,bmn} for more details of the derivation).  

For the double-well potential of the type Eq.(\ref{doublewell}), 
the saddle-point density is given in terms of two-cut eigenvalue
distribution \footnote{%
We assume $g^2_{\rm eff} < T_0/2$ so that the parameter $\lambda_\pm$'s are 
real-valued. At $ g^2_{\rm eff}= T_0/2$, $\lambda_-$ vanishes and the 
two cuts merge into one cut, signifying spill-over of eigenvalues from 
each potential well into the other.}:
\begin{eqnarray}
\rho_{{\rm s}} (\lambda) \,\, = \left\{
\begin{array}{ccc}
{1 \over 2} g^{-2}_{\rm eff} \, \sqrt{\lambda^2 \Big(\lambda^2 -\lambda_-^2
\Big) \Big(\lambda_+^2 -\lambda^2\Big) } & \qquad {\rm for} \qquad & 
\lambda \in (-\lambda_+,-\lambda_-) \cup (\lambda_-, \lambda_+), 
\\
&  &  \\ 
\qquad 0 \qquad & {\rm otherwise}. & 
\end{array}
\right.  \label{distrn}
\end{eqnarray}
Here, 
\begin{eqnarray}
\lambda_- = \sqrt{\Big(\T_0 - 2 g^2_{\rm eff} \Big)}, \qquad\quad 
\lambda_+ = \sqrt{\Big(\T_0 + 2 g^2_{\rm eff} \Big)}.
\nonumber
\end{eqnarray}
As explained above, Eq.\eq{distrn} is the leading large-$N$, large-$\theta$
value of the quantum corrected density function, in the planar limit (b),
corresponding to the classical $(N_1, N_2)$ instanton of Sec 2.2
for $N_1= N_2$.

Eq.\eq{distrn} is clearly different from the 
classical (GMS) value (Eq.\eq{densitysolution} with $N_1= N_2$).
But from what we have discussed above, 
we expect to recover the classical (GMS) value
in the weak `t Hooft coupling limit, $ g_{\rm eff} \rightarrow 0$.
This is indeed what happens. In this limit,
the eigenvalue density, Eq.(\ref{distrn}), reduces to 
\bea
\rho_{\rm s}(\lambda) \,\, \longrightarrow \,\, \rho_{\rm classical}(\lambda) 
= {\frac{1 }{2}} \delta (\lambda -
\T_0) + {\frac{1 }{2}} \delta (\lambda + \T_0).  
\nonumber
\eea
This is identical with Eq.(\ref{densitysolution}) for 
\begin{eqnarray}
N_1 = N_2 = {\frac{N }{2}} \qquad {\rm equivalently} \qquad n_1 = n_2 = {%
\frac{1 }{2}}.  \nonumber
\end{eqnarray}

\begin{figure}[htb]
   \vspace{0.5cm}
\centerline{
   {\epsfxsize=12cm
    \epsfysize=5cm
    \epsffile{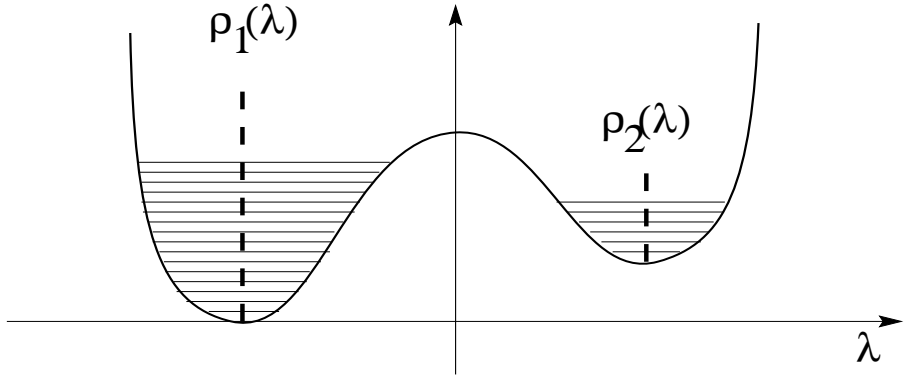}}
}
\caption{\sl Eigenvalue distributions for noncommutative vacua and 
instantons. Classically, the eigenvalues are piled up to delta-function
type distribution, as depicted as the dashed line. Quantum mechanically,
the eigenvalues repel each other and are spread over, as depicted by
the horizontal lines.}
\label{regularization}
\end{figure}

It is worth mentioning that, for the  scaling (b), the classical
limit $g_{\rm eff} \rightarrow 0$ of the planar saddle-point configuration
is not necessarily the same as the classical regime (a). 
As the result Eq.\eq{distrn} for the double-well potential 
exemplifies, the `classical limit' $ g_{\rm eff}
\rightarrow 0$ yields, out of $N$ possible classical instantons of type 
$(N_1, N_2)$, the one with $N_1 = N_2 = N/2$ singled out.

In fact, it is possible to visualize the quantum instantons as 
{\sl non-interacting} quantum vacua, localized at $\lambda_-$ and $\lambda_+$,
respectively. Consider the situation that the two potential wells are 
widely separated and contain $N_1, N_2$ eigenvalues, respectively. Then,
the partition function reads
\bea
\Z[\theta, V; n_1, n_2]
= \int_{-\infty}^\Lambda \! \prod_{k = 1}^{N_1} 
\d \lambda_k \int_{\Lambda}^{+\infty} \!\!\prod_{\ell = N_1}^N 
\d \lambda_\ell \, \prod_{1 \le k < \ell \le N} (\lambda_k - \lambda_\ell)^2
\, \exp \Big( - \S_{\rm NC} [\theta, V; N_1, N_2] \Big).
\nonumber
\eea
Here, $\Lambda$ denotes a suitably chosen, midpoint `cutoff' 
value of the eigenvalue between the two cuts. 
In fact, the above partition function is expressible as a
matrix integral over two separate matrices: ${\bf T}_1$  of size 
$(N_1 \times N_1)$ and ${\bf T}_2$ of size $(N_2 \times N_2)$, whose
eigenvalues are restricted to be less than or larger than $\Lambda$, 
respectively. One easily finds that:
\bea
\Z[\theta, V; n_1, n_2]
= \int [\d {\bf T}_1]_{N_1} \int [\d {\bf T}_2]_{N_2}
\,
\exp \Big( - \S_{\rm NC} [\theta, V; N_1, N_2] \Big),
\nonumber
\eea
where
\bea
\S_{\rm NC} &=& {\rm } \theta \, \left[ \tr V({\bf T}_1)
+ \tr V({\bf T}_2) \right] \nonumber \\
&+& \left[ 2 \tr \left( \ln ({\bf T}_1 \otimes \identity
- \identity \otimes {\bf T}_2) \right) + \cdots \right],
\label{effnc}
\eea
in which the ellipses denote gradient corrections. 
The above effective two-matrix integral is well-defined in the 
large-$N$, large-$\theta$ limit. Evidently, at leading-order in 
$(1/\theta)$-expansion, the matrix intergral is factorized
into two disjoint one-matrix integrals, except that the eigenvalues
are bounded from above and below, respectively. 
The saddle-point configuration is described precisely by the
above solution. The error involed in ellipses in Eq.(\ref{effnc}) is of 
order $e^{ - {\cal O}(N)}$, due to tunnelling effect, 
and hence is completely negligible in the continuum limit. 

\subsection{Quantum Corrections}
The central observation in the foregoing discussion was that the quantum
effects drive the eigenvalues to repel each other --- dramatic change
when compared to the situation at classical level. To demonstrate how
striking the quantum effects are, let us compute the `quantum'
Euclidean action and compare its ground-state value with that of classical
action. The second-order perturbation theory asserts that, around the
ground-state, quantum corrections to physical quantities are typically
negative. Thus, one would expect that, once the quantum corrections
are taken into account, the Euclidean action gets lowered. In quantum
NCFT, quite to the contrary, we will find that the quantum effects
{\sl increase} the Euclidean action! This has to do with the fact that
the repulsion among eigenvalues is a purely quantum-mechanical effect,
not present at classical level at all.

Begin with the effective action of the eigenvalue density field, 
$\rho(\lambda)$, Eq.(\ref{collaction}). The saddle-point configuration is
governed by solutions of Eq.(\ref{saddleeqn}). We shall be taking a generic 
condition that the classical potential $V(\lambda)$ is a concave function of 
$\lambda$ with a global minimum at $\lambda = \lambda_{\rm s}$ and denote
$V(\lambda_{\rm s}) = V_0$. Evidently, $V(\lambda) \ge V_0$ for all $\lambda$. 
Multiplying Eq.(\ref{saddleeqn}) with $\rho_{\rm s} (\lambda)$ and then 
integrating over $\lambda$, we obtain
\bea
N^2 \left[ g^{-2}_{\rm eff} 
\int_{\cal D} \d \lambda \rho_{\rm s} (\lambda) V(\lambda)
- 2 \int \d \mu \d \lambda \, \rho_{\rm s}(\mu) 
\ln \vert \lambda - \mu \vert \rho_{\rm s} (\lambda) 
\right]
= N\theta E,
\nonumber 
\eea
where we have used the normalization condition, $\int \d \lambda \,
\rho_{\rm s} (\lambda) = 1$ and $E$ is the first-integral of motion. 
Using this relation, the quantum Euclidean action Eq.(\ref{collaction})
is re-expressible as:
\bea
S_{\rm eff}[\rho_{\rm s}]
&=& N^2 \left[
g^{-2}_{\rm eff}
\int_{\cal D} \d \lambda \rho_{\rm s} (\lambda) V(\lambda)
- \int_{\cal D} \d \lambda \int_{\cal D} \d \mu \, 
\rho_{\rm s} (\lambda) \ln \vert \lambda - \mu \vert \rho_{\rm s} (\mu)
\right]
\nonumber \\
&=& N \theta E + {1 \over 2} N \theta \left( 
\int \d \lambda \rho_{\rm s} (\lambda) V(\lambda) - E \right).
\label{saddleaction}
\eea
The first-integral of motion $E$ is fixed uniquely to $E = N {\rm } \theta
\, V_0$ by demanding that, in the weak `t Hooft coupling limit, 
$g_{\rm eff} \rightarrow 0$, the saddle-point value of the quantum Euclidean
action Eq.(\ref{saddleaction}) reduces to that of the classical Euclidean
action, $S_{\rm classical} = N {\rm } \theta \, V_0 $.  
Thus, one readily finds that the second term in Eq.(\ref{saddleaction}) 
amounts to change of the Euclidean action due to quantum effects. 

Let us now evaluate the second term in Eq.(\ref{saddleaction}), the quantum 
correction to the Euclidean action. 
First of all, from the expression, whether the correction is negative -- as 
the second-order perturbation theory suggests -- or not is easily analyzable. 
Classically, the $N$ species of eigenvalues were all sitting at a single
point $\lambda = \lambda_{\rm s}$, but, once the quantum effects are taken 
into account, they will repel each other and form a domain, denoted in
Eq.(\ref{saddleaction}) as ${\cal D}$,
of eigenvalue distribution around the point $\lambda = \lambda_{\rm s}$.
Take a generic point $ \lambda$ inside ${\cal D} $. As $V(\lambda) \ge V_0$  
by the definition of $\lambda_{\rm s}$ and $\rho_{\rm s}(\lambda)$ is 
distributed over ${\cal D}$, it follows immediately that 
\bea
\Delta E &:=& \left( S_{\rm eff} - N \theta E \right)
\nonumber \\
&=& {1 \over 2} 
N {\rm } \theta \left( \int_{\cal D} \d \lambda \, \rho_{\rm s} (\lambda)
V(\lambda) - E \right) 
\nonumber \\
&=& {1 \over 2}
N {\rm } \theta \int_{\cal D} \d \lambda \, 
\rho_{\rm s} (\lambda) \Big( V(\lambda) - V_0 \Big)
\label{reexpress}\\
&\ge& 0.
\nonumber
\eea 
This proves that the quantum correction in Eq.(\ref{saddleaction}) is 
positive, in contrast to what one expects from the second-order perturbation
theory. Evidently, the reason has to do with eigenvalue repulsion --- 
classically invisible but quantum-mechanically generated effect.
The repulsion gives rise to a positive `pressure', resulting in increase of 
the Euclidean action.

We now compute the increment of the Euclidean action explicitly. 
We will take, for simplicity, $V(\lambda) = {1 \over 2} \lambda^2$
--- an approximation applicable, at leading-order, for each cut of a generic concave potential, 
according to the result of Eq.(\ref{effnc}). 
Utilizing Eq.(\ref{kernel}) (see also \cite{bipz}), 
it is straightforward to compute $\rho_{\rm s}(\lambda)$. We find
\bea
\rho_{\rm s}(\lambda) = {1 \over 2\pi} \sqrt{4 - \lambda^2}
\,\,\qquad {\rm for} \qquad -2 \le \lambda \le +2.
\label{saddlerho}
\eea
Substituting Eq.(\ref{saddlerho}) into Eq.(\ref{reexpress}), we obtain 
(recall that here the classical energy is normalized as $N \theta E=0$)
\bea
\Delta E=  S_{\rm eff} [\rho_{\rm s}]
\nonumber
&=& N {\rm } \theta \int_0^2 \d \lambda {1 \over 2 \pi} \sqrt{4 - \lambda^2}
\cdot {1 \over 2}  \lambda^2 \nonumber \\
&=&
\frac{1}{4} N \theta.
\nonumber
\eea
Thus, the correction is  of order $N^2$ in the
planar-phase (b) of Eq.\eq{3limits1}.

We conclude this section by mentioning that quantum corrections in the
disordered phase cannot be calculated by the above procedure, as the large-$N$ 
saddle-point is irrelevant. It is readily seen, however, that
the quantum corrections in the disordered phase will be larger. In the example
analyzed below, we will see that 
$(\Delta E )_{\rm Q} \slash (\Delta E)_{\rm planar} \to \infty$ in 
the disordered phase (c) of Eq.\eq{3limits}.

\subsection{Perturbative Manifestation of the Vandermonde Effect}
The effect of $S_{\rm measure}$ in Eq.(\ref{effaction}), being originated
from the vandermonde determinant of the functional integral measure, 
ought to be obtainable in the standard Feynman diagrammatics. How does the
effects manifest themselves? We will now show that, in the context of the 
Feynman diagrammatics in Weyl formulation, the aforementioned limits 
Eq.\eq{3limits} or Eq.\eq{3limits1} is derivable at large-$\theta$ and 
large-$L$ limit \footnote{Related remarks are also made in 
\cite{VanRaamsdonk:2001jd}, though some of the interpretations are 
contrasting to ours.}.

Begin with Feynman rules defined in the Moyal formulation by 
Eq.\eq{moyalpartition} and the potential Eq.\eq{doublewell}. 
Our objective is to see how the quantum corrections differ in the three 
scaling regimes, Eq.\eq{3limits}. In computing
the effects, we keep in mind the relations Eq.\eq{map} between the 
parameters $(L, \theta)$ of ${\tori^2_\theta}$ and the parameters 
$(N, \theta)$ of ${\real^2_\theta}$. 

Expand the action around the `right vacuum' 
$\T ({\bf x}) = \T_0  + \phi({\bf x})$:
\be
S_{\rm NC} =  \theta \int \d^2 {\bf x}  
\left[ 
V_0 + \left( \lambda_4 \T_0^2 \phi^2 + \lambda_4  \T_0 \phi^3 + {\lambda_4 
\over 4 } \phi^4 \right) + \left(- {1 \over 2\theta} 
\left(\del_{\bf x} \phi\right)^2 \right)\right]_\star.
\label{a.1}
\ee
Consider, for definiteness,  the nonplanar, one-loop contribution to the
connected two-point Green function, depicted in Fig.\eq{np}.
This diagram provides an example of IR problems in
NCFT\cite{uvir}. 

\begin{figure}[htb]
\begin{center}
\epsfxsize=5cm\leavevmode\epsfbox{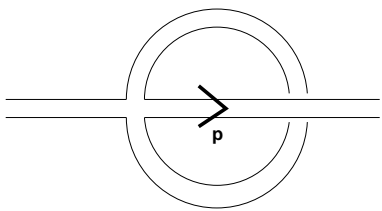}
\end{center}
\caption{\sl one-loop, non-planar contribution to two-point Green function.} 
\label{np}
\end{figure}

The contribution involves the following moduli-space integral
associated with one-loop Feynman diagram, where $D$ denotes the
spacetime dimension:
\bea
I_D= \int\limits_0^\infty \d t {1 \over t^{D/2}} \exp\left(- t m^2
	- {\Lambda_{\rm eff}^2 \over t} \right) ,
\nonumber
\eea
where $m^2 = 2$ from Eq.\eq{a.1}, ${\bf p}$ is the momentum flow
through the external line, and 
\bea
\Lambda_{\rm eff}^{-2} = \Lambda_{\rm UV}^{-2} + 2 \lambda_4 
\left( \theta \cdot {\bf p} \right)^2.
\nonumber
\eea
The result is
\bea
I_2 &=& 2 K_0\left({2m} \slash {\Lambda}_{\rm eff} \right)
\rightarrow \log\left({2m} \slash {\Lambda}_{\rm eff} \right) + \ldots 
\qquad {\rm for} \qquad \left({2m} \slash {\Lambda_{\rm eff}} \right) \ll 1
\nonumber \\
&\rightarrow& 2 K_0 \left(2m \vert \theta {\bf p}\vert \right) 
\qquad \qquad \qquad \qquad \qquad {\rm for} \qquad {\bf p} = {\rm finite}, 
\quad \Lambda \rightarrow \infty
\nonumber \\
&\sim & 2 K_0\left(2m \theta/L \right) 
\qquad \qquad \qquad \qquad \qquad 
{\rm for} \qquad \vert {\bf p} \vert \sim 1/L.
\nonumber
\eea
Using the relation $L= \sqrt{\theta N}$ on $\tori^2_\theta$, 
we finally obtain 
\be
I_2= 2 K_0 \left({2m \over g_{\rm eff}} \right).
\nonumber
\ee
Thus, $I_2 = \infty$ in the disordered phase, $I_2 = $ finite in the planar phase, 
and $I_2=0$ in the GMS phase. This is exactly as we would predict on the 
basis of our earlier discussion of the behaviour of the quantum
effective action in the limits Eq.\eq{3limits}, namely that
the GMS solution remains stable in the limit (a),
has a finite correction in the planar limit (b), and is 
completely destabilized in the limit (c), where the measure term becomes 
infinitely large compared to the classical term in the action.
 
\section{Effect of the Gradient Term}
The foregoing discussion was largely based on keeping only the leading order 
term, ${\cal L}_0$ in Eq.(\ref{twoterms}), at large ${\rm }\theta$ limit. 
While the gradient-term ${\cal L}_{-1}$ is sub-leading order in 
$(1/{\rm }\theta)$-expansion, as noted below Eq.(\ref{uinfinity}),
it breaks the U($\infty$) symmetry explicitly --- a point which ought to 
be concerned for its consequential effects to the results we have 
obtained in the previous subsections. In particular, as the dramatic
quantum effects we have deduced are largely based on ${\cal L}_0$-term
and U($\infty$) symmetry therein, one might suspect that the term 
${\cal L}_{-1}$, being part of the classical action, would render a sizable 
symmetry breaking effect. This is because the size of the gradient-term 
is given by
\bea
\S_{-1} = \int \d^2 {\bf x} {\cal L}_{-1} = \int \d^2 {\bf x} 
\left(-{1 \over 2} [{\bf x}, \T]_\star^2 \right)
\quad \sim \quad {\cal O}(N^2).
\nonumber
\eea
Fortuitously, as we will show in this section, the gradient effect turns
out to be of order ${\cal O}(N^2 g_{\rm eff}^2)$, viz. scales further 
by a factor of the `t Hooft coupling, $g^2_{\rm eff}$. The scaling is not 
universally valid, but only for $g_{\rm eff} < g^{\rm c}_{\rm eff}$ for some 
finite $g^c_{\rm eff}$, as is inferred from the large-$N$  phase 
transition \cite{gww}. As we are 
interested in the weak `t Hooft coupling regime, $g_{\rm eff} \ll 1$,
the above counting holds valid. In particular, it implies that the measure 
effect, whose size is of order ${\cal O}(N^2)$, outweighs the gradient
effect. Thus, in the weak `t Hooft coupling regime, one can utilize the 
U$(\infty)$ symmetry, and recast the NCFTs literally as the $N \rightarrow
\infty$ limit of the matrix model studied in \cite{bipz}. 
\subsection{Perturbative Estimates}
We will begin with, utilizing the Weyl formulation of the NCFT, computation 
of leading-order perturbative corrections. 
For this purpose, we regularize the theory so that fields are defined on 
$N$-dimensional Hilbert space, ${\cal H}_{\theta}[N]$, spanned by
Span[$| n \rangle, n=0, \ldots, N-1$], where
\bea
| n \rangle = \frac{a^{\dagger n}}{\sqrt{n!}} |
0 \rangle \qquad {\rm and} \qquad 
\frac{1}{\sqrt 2}[\widehat{{\bf x}^1} \pm i \widehat{{\bf x}^2}] 
\equiv ({\bf a}, {\bf a}^\dagger) .
\nonumber
\eea
Taking the potential as Eq.(\ref{doublewell}) and expanding around 
${\bf T} = 0$, the NCFT partition function Eq.(\ref{weylpartition}) 
is given by \footnote{Actually, in Eq.\eq{doublewell}, ${\bf T}=0$ is an 
unstable point. One might alternatively expand the potential around stable 
vacua, ${\bf T}= \pm \T_0 \identity$. This would give rise to an additional
cubic interaction, but it turns out that the conclusion based on 
Eq.\eq{quartic} remains unchanged.}
\bea
\Z_N = \int [\d{\bf T}]_N \exp \left( - \S_{\rm NC}[\theta; {\bf T}] 
\right), \qquad {\rm where} \qquad
\S_{\rm NC} = \Big( \S_{\rm cl} + \S_{\rm 0P} + \S_{\rm 0V} \Big) + \S_{-1}.
\nonumber
\eea
Here,
\bea
\S_{\rm cl} &=& \theta \tr V(0) \,\, \qquad \qquad {\rm where} \qquad
 V(0) = V_0 + {\lambda_4 \over 4} \T_0^4 \nn 
\S_{\rm 0P} &=& \theta \tr \left( {m^2 \over 2} {\bf T}^2\right)
\qquad {\rm where} \qquad m^2 = {\lambda_4 \over 2} \T^2_0
\nn
\S_{\rm 0V} &=& \theta \tr \left( {\lambda_4 \over 4}  {\bf T}^4 \right)
\nn
\S_{-1} &=& \tr \left( - \frac12 \left[\widehat{\bf x}, {\bf T} \right]^2
\right).
\label{quartic} 
\eea
\subsection{Leading-Order Corrections}
The leading-order, quantum corrections to the free energy are obtained as
\bea
\Big< \S({\bf T}) \Big>_{\rm 0P} = 
\frac{1}{\Z_{\rm 0P}} \int [\d {\bf T}] \S({\bf T}) \exp \Big(-\S_{\rm 0P} \Big)
= \S_{\rm cl} + \Delta \S_{\rm 0P} + 
\Delta \S_{\rm 0V} + \Delta \S_{-1} + \ldots 
\label{energy-expansion}
\eea
where 
$\Z_{\rm 0P} := \int [\d {\bf T}] \exp \Big(- \S_{\rm 0P} \Big)$ and 
\bea
\S_{\rm cl} &\equiv& \theta {\rm Tr} V(0) \sim \theta N V(0), 
\qquad \qquad \quad
\Delta \S_{\rm 0P} \equiv \theta m^2 \ \Big<  \tr {\bf T}^2 \Big>_{\rm 0P} 
\sim N^2
\nn
\Delta \S_{\rm 0V} & \equiv & \Big< \S_{\rm 0V} \Big>_{\rm 0P} = 
\theta \lambda_4 \cdot \frac{N^3}{(\theta m^2)^2}, 
\qquad \quad
\Delta \S_{-1} \equiv \Big< \S_{-1} \Big>_{\rm 0P} = 
\frac{N \cdot N^2}{\theta m^2}.  
\label{energy-expansion-value}
\eea

\begin{figure}[htb]
   \vspace{0.5cm}
\centerline{
   {\epsfxsize=8.5cm
   \epsfysize=6cm
   \epsffile{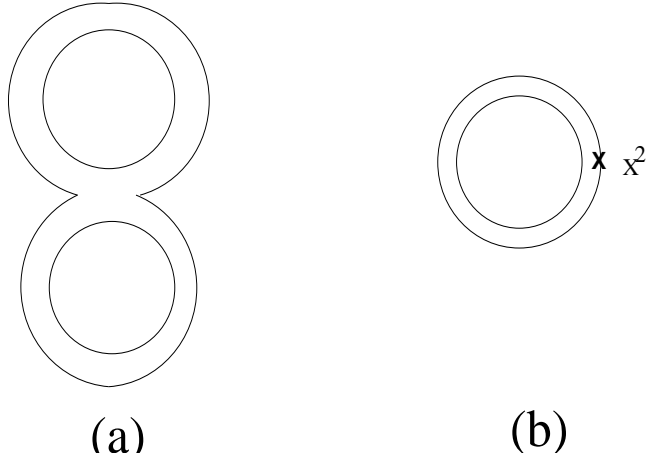}}
}
\caption{\sl Feynman diagrams for leading-order quantum corrections due to 
(a) potential $S_{0V}$, and (b) the gradient term $S_{-1}$. }
\label{derivative}
\end{figure}
Diagrammatically, $\Delta \S_{\rm 0P}$ originates from the one-loop vacuum 
diagram, while $\Delta \S_{\rm 0V}$ and $\Delta \S_{-1}$ are from the 
diagrams (a) and (b) in Fig.\eq{derivative}, respectively. Note that the 
propagator is given by $\Big< {\bf T} {\bf T} \Big>_{\rm 0P} 
\sim 1/(\theta m^2)$.  
Thus, the diagrams are evaluated as follows. For diagram (a), the contribution
equals to ({\rm vertex}) $[\theta  \lambda_4] \times$ (two~ propagators) 
$[ 1/ (\theta m^2)^2 ] \times$ (three `color' loops) $[ N^3] $. 
In evaluating $\Big< \S_{-1} \Big>_{\rm 0P}$, two terms will contribute: 
$\tr \left( {\bf T} \hat{\bf x}_a {\bf T} \hat{\bf x}_a \right)$ and 
$\tr \left( {\bf T}^2 \hat{\bf x}^2_a \right)$. 
As $\tr \hat{\bf x}_a = 0$, only the latter will 
contribute, and is  given by the Feynman diagram (b) in Fig.\eq{derivative}. 
There, ${\rm X}^2$ on the outer colour loop refers to the insertion of the 
$\tr \left( \widehat{\bf x}_a^2 \right) \sim
\sum_{n=0}^{N-1} \, n  = {\cal O}(N^2)$. Hence, for diagram (b), the
contribution equals to (color loop) $[N] \times$  (color loop with 
${\rm X}^2$ insertion) $ [N^2] \times$ (one propagator) [$ 1/(\theta m^2)]$. 

To proceed further, introduce the following rescaled parameters: 
\bea
\overline{\lambda_4} := \frac{\lambda_4}{m^2}, 
\qquad
\overline{V}(0) := \frac{V(0)}{m^2},
\qquad
\overline{\theta} := \frac{m^2 \theta}{N} = {m^2 \over g^2_{\rm eff}}.
\nonumber
\eea
Making, in Eq.\eq{quartic}, a change of the variable 
$\theta m^2 {\bf T}^2 = {\bf M}^2$ and bringing the quadratic term into
a canonical normalization, we have 
\bea
\S_{\rm NC} = N^2 \overline{\theta} \overline{V}(0) 
+ \tr \left[ \ {1 \over 2} {\bf M}^2 + 
{1 \over N} \frac{\overline{\lambda_4}}{\overline\theta}
\left( {1 \over 4} {\bf M}^4 \right) 
+ {1 \over N} \frac{1}{\overline\theta}\left( -{1 \over 2}
[\widehat{\bf x}_a, {\bf M}]^2 \right) \right],
\label{rescaledaction}
\eea
with which the partition function Eq.\eq{matrixpartition} can be defined.
The Eq.(\ref{rescaledaction}) 
reveals that the effective coupling of the potential term
is $(\overline{\lambda_4} \slash \overline{\theta})$ and that of the
gradient term is $\frac{1}{\overline\theta}$. For the perturbation theory
to make sense, one will need these couplings to be small enough. 
We now ask if there is a range of parameters satisfying
this restriction as well as the condition that the gradient 
terms are suppressed compared to the potential term.
There indeed does exist such a region in the space of the rescaled 
parameters, viz.
\be
\overline \theta \, \gg \, \overline{\lambda_4} \, \gg \, 1.
\label{region} 
\ee
We will now explicitly verify that, in this weak coupling regime, the gradient 
term is suppressed, at least at leading order in the perturbation theory.
In terms of the rescaled parameters, the estimates 
Eq.\eq{energy-expansion-value} are re-expressible as 
\bea
\S_{\rm cl} = N^2 \overline{\theta} \overline{V}(0),  \qquad 
\Delta \S_{\rm 0P} = N^2, \qquad 
\Delta \S_{\rm 0V} = N^2 \left( \frac{\overline{\lambda_4}}{\overline\theta}
\right), 
\qquad \Delta \S_{-1} =  N^2 \left( \frac{1}{\overline\theta} \right).
\label{energy-expansion-value-1}
\eea
We thus realize that, in the 't Hooft's large-$N$ limit, all the terms are of 
order ${\cal O}(N^2)$, and hence are planar.
However, the weak coupling limit ensures that the leading-order
corrections are hierarchically ordered
\bea
\S_{\rm cl} \, \gg \, \Delta \S_0 \, \gg \, \Delta \S_1 \, \gg \,
\Delta \S_{-1}.
\label{hierarchy}
\eea
Hence, we conclude that, under Eq.\eq{region}, the gradient term 
$\S_{-1}$ is indeed suppressed compared to the potential term.

\subsection{Higher-Order Corrections}
To ensure that the scaling limit Eq.(\ref{region}) is sufficient for dropping
the gradient terms at least perturbatively, we now 
evaluate next-to-leading order
corrections. They arise from the second-order expansion of the partition
function, and is given by the connected vacuum diagrams, 
Fig.7:
\begin{figure}[htb]
   \vspace{0.5cm}
\centerline{
   {\epsfxsize=9cm
   \epsfysize=6cm
   \epsffile{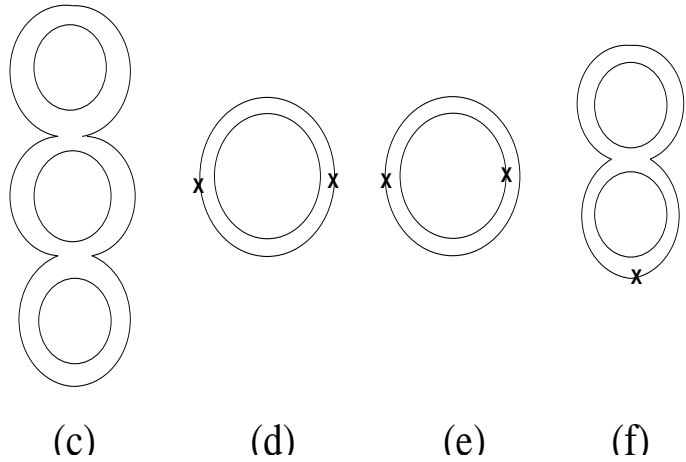}}
}
\caption{\sl Feynman diagrams for higher-order quantum corrections due to 
(c) potential $(S_{\rm 0V})^2$, and (d),(e) the gradient terms $(S_{-1})^2$,
and (f) the cross term $(S_{\rm 0V} S_{-1})$. }
\label{derivative1}
\end{figure}

\bea
\Big< {1 \over 2!} \S^2 \Big>^{\rm conn}_{\rm 0P} = {1 \over \Z_{\rm 0P}}
\int\limits_{\rm connected} [\d {\bf M}] 
{1 \over 2!} \left(\S_{\rm 0V} + \S_{-1} \right)^2
\exp \left( - \S_{\rm 0P} \right).
\nonumber
\eea
Dropping again `dimensionless' numerical factors of ${\cal O}(1)$, we
obtain the corrections as
\bea
{\rm diagram~ (c)} &\sim& 
\,\, \,\, \left(\theta \lambda_4 \right)^2  \frac{N^4}{(\theta m^2)^3} 
\,\, = \, \, N^2 
\left( {\overline{\lambda_4} \over \overline{\theta}} \right)^2,
\nn
{\rm diagram~ (d)} &\sim & \,\, \quad \,\,\,
{N \cdot N^3 \over (\theta m^2)^2} \qquad = N^2 
\left({1 \over \overline{\theta}} \right)^2, 
\nn
{\rm diagram~ (e)} &\sim & \,\, \quad \, 
{N^2 \cdot N^2 \over (\theta m^2)^2} \qquad
= N^2 \left( {1 \over \overline{\theta}} \right)^2,
\nn
{\rm diagram~ (f)} &\sim&
(\theta \lambda_4) \frac{N \cdot N \cdot N^2}{(\theta m^2)^3}
= N^2 \left( \frac{\overline{\lambda_4}}{\overline\theta} \right). 
\label{higher}
\eea
Evidently, an insertion of the gradient term $\S_{-1}$ is accompanied by 
an extra factor of $\overline{\lambda}_4 \slash \overline{\theta}$.
In the scaling limit Eq.\eq{region}, the factor is small enough. 
We thus conclude that, by taking the scaling limit Eq.\eq{region}, 
effect of the gradient terms can be made hierarchically small compared 
to the vandermonde effect.
\subsection{Nonperturbative Estimate}
We will now make use of Feynman's variational method \cite{feynman, sakita}, 
and prove 
nonperturbatively that the scaling limit Eq.\eq{region} ensures subdominance
of the gradient terms. From Eq.\eq{analogy} expressed in terms of the
rescaled action Eq.\eq{rescaledaction},  
\bea
\Z_N &:=& \exp \Big( - F_{\rm exact} \Big)
\nonumber \\
&=& \int [\d {\bf M}] \exp \Big( - \S_0 \Big) \exp \Big( - \S_{-1} \Big)
\nonumber \\
&=& \int [\d {\bf M}] \exp \Big( - \S_0 \Big) \Big< \exp \Big(- \S_{-1} \Big) 
\Big>_0,
\nonumber
\eea
where $F_{\rm exact}$ refers to the exact free-energy, 
$\S_0 = (\S_{\rm 0P} + \S_{\rm 0V})$, and 
\bea
\Big< \cdots \Big>_0 := \int [\d {\bf M}] \cdots \exp(-\S_0) 
\Big\slash \int [\d {\bf M}] \exp(-\S_0).
\nonumber
\eea 
Applying Jensen's inequality, we have
\bea
\Z_N \ge \int \prod_{a=1}^N [\d {\bf M}]
\exp \Big( -\S_0 \Big) \, \exp \left( - \Big< \S_{-1} \Big>_0 \right).
\nonumber
\eea
Thus, we find a variational estimate to the upper-bound of the exact
free-energy:
\bea
F_{\rm exact} \le F_0 + \Delta F \qquad
\nonumber
\eea
where
\bea 
F_0 = - \ln \int [\d {\bf M}] \exp \Big( - \S_0 \Big) \qquad
{\rm and} \qquad \Delta F :=  \Big< \S_{-1} \Big>_0.
\label{correction}
\eea

The quantity $\Delta F$ can be evaluated explicitly by utilizing the
well-known formulae \cite{creutz}:
\bea
\Big< {\bf M}_{kl} {\bf M}_{mn} \Big>_0 
=  C_1 \delta_{kl} \delta_{mn} + C_2 \delta_{kn} \delta_{lm},
\label{correlator}
\eea
where 
\bea
C_1= { \Big< (\tr{\bf M})^2\Big>_0 \over (N^2 - 1)} 
- { \Big<\tr {\bf M}^2\Big>_0 \over N (N^2 - 1)} 
\qquad {\rm and}\qquad  
C_2= {\Big<\tr {\bf M}^2 \Big>_0 \over (N^2 - 1)} 
- { \Big<(\tr {\bf M})^2 \Big>_0 \over N(N^2- 1)}.
\nonumber
\eea
The Eq.\eq{correlator} can be proved from the ${\rm U_L}(N) \times {\rm U_R}(N)$ invariance
of both the action $\S_0$ and the integral measure $[\d {\bf M}]$.
Hence, the correction $\Delta F$ in Eq.\eq{correction} is computed as
\bea
\Delta F &=& \Big< {1 \over \overline{\theta} }{1 \over N} 
\tr\left( - {1 \over 2} [\widehat{\bf x}^a, {\bf M}]^2 \right) \Big> 
\nonumber \\
\nonumber \\
&=& {1 \over \overline{\theta}}
\left[ -\left(\widehat{\bf x}^a_{nk} \widehat{\bf x}^a_{lm} \right)
{1 \over N} \Big< {\bf M}_{kl} {\bf M}_{mn} \Big>_0 + 
\left(\widehat{\bf x}^a \widehat{\bf x}^a \right)_{nk} 
{1 \over N} \Big<({\bf M}^2)_{kn} \Big>_0 \right] 
\nonumber \\
\nonumber \\
&=& \left[ {1 \over N (N^2 - 1)}
\left(N \Big<\tr {\bf M}^2 \Big>_0 - \Big<(\tr {\bf M})^2 \Big>_0 \right)
\right] {1 \over \overline{\theta}}
\tr\Big(\widehat{\bf x}^a \widehat{\bf x}^a \Big),
\nonumber
\eea
where, in the last equality, we have used the fact $\tr {\bf x}^a = 0$.
Evidently, from the last expression, 
as $\Big< \tr {\bf M}^2 \Big>$ and $\Big< (\tr {\bf M})^2 \Big>$ scales with
$N$ as ${\cal O}(N^2)$, the coefficients inside the square bracket is of 
order ${\cal O}(1)$. Hence, $\Delta F$ is proportional to $\tr\Big(\widehat{\bf x}^a \widehat{\bf x}^a \Big) \slash
\overline{\theta}$, and is of order ${\cal O}(N^2 \slash \overline{\theta})$. 
As the $\S_{\rm classial}$ and $\Big< \S_0 \Big>$ scale as 
${\cal O}( N^2 \overline{\theta})$ and ${\cal O}(N^2)$, respectively, 
the above estimate indeed shows that the gradient term contribution is bounded 
from above to a value suppressed by powers of $1/\overline{\theta}$. 
This completes the proof that Eq.\eq{hierarchy} holds at nonperturbative
level.  
\subsection{Remarks on Gradients in Gauge Theories}
In case the NCFT is promoted to gauge theories, the situation becomes even 
more favorable. 
In this section, we have also restricted our investigation to 
NCFTs consisting only of scalar fields --- corresponding to the level-zero 
truncation in the context of open string field theory. 
Once the gauge field is coupled, the action is schematically given as
\bea
S_{\rm NC} = \int_{\real^2} \d^2 {\bf y} 
\left[ {1 \over 4} F_{mn} \star_\theta F_{mn} + 
{1 \over 2} D_m (A) \T \star_\theta D_m(A) \T 
+ V_{\star_\theta} (\T) + \cdots \right],
\nonumber
\eea 
corresponding to truncation of the open string field theory at level-one.
Here, $D_m(A) \T := \partial_m \T + \left[ A_m, \T \right]_\star$, and, via 
the Weyl-Moyal map, it is expressible as being proportional to 
$[{\bf Y}_m,\T]$, where ${\bf Y}^m := {\bf y}^m + \theta^{mn} A_n({\bf y})$. 
A crucial observation for
the present discussion is that, as \cite{gms2} have pointed out, 
${\bf Y}^m = 0$ in classical vacua at 
any nonzero value of ${\rm }\theta$. Because of this, the tachyon gradient 
term, $[{\bf Y}_m, {\bf T}]^2$, drops out of the Euclidean action completely.
Moreover, this nullification takes place for finite value of 
${\rm }\theta$.

\section{D = (2+1) Noncommutative Field Theories}
We next turn our attention to noncommutative field theories in 
$(2+1)$-dimensional spacetime. As in the previous sections, our main 
motivation concerning these theories would be that theses theories describe 
tachyon dynamics on an unstable D2-brane, either in bosonic or in Type IIB 
superstring theories, at nonzero B-field background. The main
result we shall be showing is that, at low-energy, {\sl quantum} aspects of
vacua and solitons (corresponding to non-BPS D0-branes) are governed by 
{\sl quantum mechanics} of a $(0+1)$-dimensional Hermitian matrix model. 
Moreover, we again find that the continuum and semiclassical limit is governed 
by large-$N$, large-$\theta$ limit.
Most of the discussions are closely parallel to the two-dimensional case
of the previous section. Nevertheless, for the sake of readers, we will repeat 
those parts relevant for foregoing discussions.
\subsection{Classical Theory}
Begin with noncommutative $(2+1)$-dimensional spacetime 
$\real^{2,1}_\theta$, whose coordinates are $(t,{\bf y})$
 and `spacelike' 
noncommutativity are $\theta^{ab}$:
\[
\left[ y^a,y^b\right] =i\theta^{ab} \qquad {\rm and} \qquad 
\left[t, y^a \right] = 0, \qquad (a,b=1,2).
\]
Take a field theory on $\real^{2,1}_\theta$, consisting of
a scalar field $\T(t, {\bf y})$ with self-interaction potential $V(\T)$. 
The Seiberg-Witten map enables us to map the theory into to a noncommutative 
field theory on $\real^{2,1}$, whose action is given by: 
\bea
S_{\rm NC}[\theta; {\rm V}]=\int_{\real^{2,1}}  \d t \d^2{\bf y}\,
\left[ {\frac 12}\partial _t \T \star_\theta \partial_t \T-{%
\frac 12}\partial _{{\bf y}} \T \star_\theta \partial_{{\bf y}} \T
-V_{\star }(\T) \right].
\label{3daction}
\eea
The noncommutativity $\theta^{ab}$ is encoded into the $\star_\theta $-product, 
defined as before, Eq.(\ref{moyalprod}). We are again interested in the 
large noncommutativity limit, ${\rm } \theta \rightarrow \infty $. 
Rescale the spatial coordinates, ${\bf y} \rightarrow {\bf x}$, the same way 
as in Eq.(2.4), and expand  the action Eq.(\ref{3daction}) in powers of 
$(1/{\rm }\theta)$: 
\bea
S_{\rm NC}[\theta; V]
= {\rm } \theta \int_{\real^{2,1}} \d t \d^2 {\bf y}\,
\left[ {\cal L}_0+\frac1{{\rm } \theta }{\cal L}_{-1}+ \cdots\right],
\label{expanded3daction}
\eea
where 
\bea
{\cal L}_0=
{1 \over 2} \left( (\partial_t \T)^2 - V_{\overline{\star }}(\T) \right)
\qquad {\rm and}\qquad 
{\cal L}_{-1}=-{\frac 12}(\partial_{{\bf x}} \T)^2. 
\nonumber
\eea
Again, at large noncommutativity, $(1/ {\rm } \theta) \rightarrow
\infty$, the gradient-term ${\cal L}_{-1}$ drops out. 

The aforementioned Weyl-Moyal map,
\bea
\T({\bf x}, t) = \int_{\widetilde{\real^2}} {\d^2 {\bf k} \over (2 \pi)^2}
\tr_{{\cal H}} 
\left( e^{ i {\bf k} \cdot \widehat{\bf x}} {\bf T}(t) \right) 
e^{ - i {\bf k} \cdot {\bf x}},
\nonumber
\eea
then permits us to re-express the (2+1)-dimensional NCFT 
Eq.(\ref{expanded3daction}) as a one-dimensional Hermitian matrix model:
\begin{eqnarray}
\S_{\rm NC}[\theta; V] = {\rm } \theta \, \int \! \d t \, 
\tr_{{\cal H}} 
\left[ \left({1 \over 2} (\partial_t {\bf T})^2 - V({\bf T}) \right) 
+ {1 \over {\rm } \theta} \left( + {1 \over 2} [\hat{\bf x}, {\bf T}]^2
\right) +\cdots \right].  
\label{op3daction}
\end{eqnarray}
At leading order in $(1/{\rm }\theta)$, both Eq.(\ref{expanded3daction})
and Eq.(\ref{op3daction}) are invariant under the U($\infty$) symmetry group
of area-preserving diffeomorphism:
\bea
\T({\bf x}, t) \rightarrow U({\bf x}, t) \star
\T({\bf x}, t) \star U^{-1} ({\bf x}, t)
\qquad \longleftrightarrow \qquad 
{\bf T} \rightarrow {\bf U}(t) \,  {\bf T}(t) \, {\bf U}^{-1}(t).
\nonumber
\eea
The scalar field, realized as an operator field ${\bf T}(t)$ on the auxiliary
Hilbert space ${\cal H}$, is expandable as a linear combination of
one-dimensional projection operators:
\bea
{\bf T}(t) = \sum_{\ell = 1}^{{\rm dim}{\cal H}}
\lambda_{a_\ell}(t) {\bf P}_\ell,
\nonumber
\eea
where the one-dimensional projection operators ${\bf P}_\ell$'s are defined
as in Eq.(\ref{projop}) and the coefficients $\lambda_a$'s are generically
time-dependent. 

\subsection{Classical Vacua and Solitons}
Utilizing the one-dimensional projection operators ${\bf P}_\ell$'s, it is
straightforward to construct {\sl static} classical solutions, as shown first 
in \cite{gms}. Denote critical points of the potential, defined by 
$V^{\prime}(\lambda) = 0$, as 
$\lambda_0, \lambda_1, \lambda_2, \cdots$, arranged in ascending order 
of the critical point energy: $V(\lambda_0) \le V(\lambda_1) \le V(\lambda_2) 
\le \cdots$. The most general solution to the tachyon equation of motion
\bea
-\partial_t^2 {\bf T}(t) - V'({\bf T}) = 0
\nonumber
\eea
is expressible as 
\bea
{\bf T}(0) = \sum_{\ell = 1}^{{\tt dim} {\cal H}} 
\lambda_{a_\ell} {\bf P}_\ell.
\nonumber
\eea
where the coefficients $\lambda_{a_\ell}(t)$ obeys single-particle equation
of motion
\bea
- \ddot\lambda_{a_\ell}(t) - V'(\lambda_{a_\ell}(t)) = 0.
\nonumber
\eea
Evidently, the $a$-th vacuum is given by
\bea
{\bf T}_{a} = \sum_{\ell = 1}^{{\tt dim} {\cal H}} 
\lambda_a {\bf P}_\ell = \lambda_a \identity_{{\cal H}}
\qquad (a = 0, 1, 2, \cdots),
\nonumber
\eea
where $\lambda_{a_\ell}$'s take values out of the set $(\lambda_0, \lambda_1,
\cdots)$ permitting duplications. Likewise, a classical static soliton is 
given by 
\bea
{\bf T}_{\rm soliton}(t) = \sum_{\ell = 1}^{{\tt dim} {\cal H}} 
\lambda_\ell {\bf P}_\ell  
\nonumber
\eea
where the coefficients $\lambda_\ell$'s consist of {\sl at least} two 
distinct values among the critical points. For example, a static soliton of 
type $(N_a, N_b)$ is given by 
\bea
{\bf T}_{(N_a, N_b)} = \lambda_a {\bf P}_{[N_a]}  
+\lambda_b {\bf P}_{[N_b]}. 
\label{soliton}
\eea 
Not surprisingly, the soliton takes the same form as the $[N_a, N_b]$ 
instanton considered in section 2, as, even in noncommutative context, 
instantons in (2+0)-dimensional NCFT are identifiable with static configuration
of  solitons in (2+1)-dimensional NCFT. 

To exemplify this, consider again the symmetric double-well potential:
\bea
V({\bf T}) = V_0 + {\lambda_4 \over 4} \left( {\bf T}^2 - \T_0^2 \right)^2
.
\nonumber
\eea
The classical vacua are given by the linear operators
\bea
{\bf T}_{\rm vacuum} = \pm \T_0 \identity,
\nonumber
\eea
while the static $(N_1, N_2)$ soliton is given by 
\bea
{\bf T}_{N_1, N_2} = \T_0 \left( {\bf P}_{[N_1]} - {\bf P}_{[N_2]} \right).
\nonumber
\eea
The U($\infty$)-invariant collective excitations are encoded into the 
eigenvalue density field $\rho(\lambda, t)$ \cite{collective}:
\bea
\rho(\lambda, t) := {1 \over {\tt dim} {\cal H}}
\sum_{\ell=1}^{{\tt dim} {\cal H}} 
\delta \Big(\lambda - \lambda_\ell (t)  \Big).
\label{rho-def}
\eea
For example, the static $(N_1, N_2)$ soliton is then given by saddle-point 
configuration of the density field $\rho_{\rm s}(\lambda)$:
\bea
\rho_{[N_1, N_2]}(\lambda) = n_1 \delta (\lambda - \T_0)
+ n_2 \delta (\lambda + \T_0)
\qquad 
{\rm where}
\qquad
n_{1,2} = {N_{1,2} \over {\tt dim} {\cal H}}.
\label{classicaldistrn}
\eea

\subsection{Quantum Theory}
\subsubsection*{Definition}
For the definition of the theory at quantum level, we will adopt the same 
prescription as the (2+0)-dimensional case. 
Thus, in the Moyal formulation via (2+1)-dimensional NCFT, 
the \underline{regularized} partition function is defined as:
\bea
{\cal Z}_{\rm NC} [\theta, V_\star; L_1 L_2]
= \int[\d \T(t)]_{L_1, L_2} \,
\exp \Big( - S_{\rm NC} [\theta; V_\star(\T)] \Big).
\nonumber
\eea
In the Weyl formulation via (0+1)-dimensional Hermitian matrix model,
the \underline{regularized} partition function is defined as:
\bea
\Z_N[\theta, V; N] = \int [\d {\bf T}(t)]_N 
\exp \Big( - \S_{\rm NC}[\theta; V({\bf T}] \Big),
\nonumber
\eea
where the integration measure is defined as in (2+0)-dimensional NCFT:
\bea
[\d \T]_N := \prod_{-\infty < t <+\infty} \left( 
\prod_{\ell=1}^N \d \T_{\ell\ell}(t) \prod_{1 \le \ell < m \le N} 
2 \d {\rm Re} \T_{\ell m }(t) {\rm Im} \T_{\ell m}(t) \right).
\label{3dmeasure}
\eea
The Weyl-Moyal correspondence then implies that
\bea
\lim_{L_1 L_2 \rightarrow \infty}
{\cal Z}_{\rm NC} [\theta, V_\star; L_1 L_2] \quad
\equiv
\lim_{N \rightarrow \infty} \Z_N [\theta, V; N].
\nonumber
\eea

We will thus investigate the quantum effects in terms of the right-hand
side, viz. the (0+1)-dimensional Hermitian matrix model. We are interested 
in computing ground-state energy and low-energy excitations of the theory. 
>From Eq.(\ref{weylaction}) and the definition of the integration measure 
Eq.(\ref{3dmeasure}), one readily obtains the Hamiltonian as 
\bea
\H = -{1 \over 2 {\rm } \theta} \Delta_{\bf T} + {\rm } \theta \, 
\tr V({\bf T})  + \Delta \H_{\rm grad},
\nonumber
\eea
where
\bea
\Delta_{\bf T} := 
- \tr \Pi_{\bf T}^2 &=& \sum_{\ell = 1}^N {\partial^2 \over \partial 
{\bf T}_{\ell \ell}^2}
+ {1 \over 2} \sum_{1 \le \ell < m \le N} 
\left( {\partial^2 \over \partial {\rm Re} {\bf T}_{\ell m}^2}
     + {\partial^2 \over \partial {\rm Im} {\bf T}_{\ell m}^2} 
\right) \nonumber \\
\Delta \H_{\rm grad} &=& \Big( - {1 \over 2} \left[\widehat{\bf x}, {\bf T}
\right] \Big)^2.
\label{completehamiltonian}
\eea
For now, anticipating a similar power-counting suppression as in the
two-dimensional NCFTs, we will drop the gradient term $\Delta \H_{\rm grad}$, 
and justify it later in section 4.5. 
Parametrize the matrix field ${\bf T}(t)$ as
\bea
{\bf T}(t) = {\bf U}(t) \cdot {\bf T}_{\rm d} (t) \cdot {\bf U}^{-1}(t),
\label{diag}
\nonumber
\eea
where
\bea
{\bf T}_{\rm d}(t) = {\rm diag}.
\left( \lambda_1 (t), \cdots, \lambda_N(t) \right).
\nonumber
\eea
The `angular' matrix ${\bf U}(t)$ parametrizes coset space $SU(N)/{\cal W}$,
where ${\cal W}$ refers to the Weyl group, permuting the eigenvalues.
Evidently, as the Hamiltonian is invariant under the U($\infty$) 
transformation, the ground-state wave function $\Psi({\bf T})$ ought to be a 
symmetric function of the eigenvalues $\lambda_\ell$ of ${\bf T}$. 
The ground-state energy is given by: 
\bea
E_{\rm g.s.} 
= \lim_{N \rightarrow \infty}
{\rm Min}_{\Psi}
{\left< \Psi \vert \H \vert \Psi \right> \over 
\left< \Psi \vert \Psi \right>}
\nonumber
\eea
over the variational wave functions $\Psi$. Here, the matrix-elements are
\bea
\left< \Psi \vert \H \vert \Psi \right> = 
\int [\d {\bf T}] \Psi^\dagger (\lambda) \H \Psi (\lambda)
\qquad
{\rm and}
\qquad
\left< \Psi \vert \Psi \right>
= \int [\d {\bf T}] \Psi^\dagger (\lambda) \Psi (\lambda) .
\nonumber
\eea
Note that the ground-state
wave function $\Psi$ is invariant under the transformation
${\bf T}(t) \rightarrow {\bf U}(t) {\bf T}(t) {\bf U}^{-1}(t))$. 
Eliminating the `angular' variables $ {\bf U}(t)$, the matrix elements 
can be rewritten as
\bea 
\left< \Psi \vert \H \vert \Psi \right> 
&=& \int \prod_{\ell = 1}^N \d \lambda_\ell \, \Delta^2 (\lambda)
\left( {1 \over 2} \sum_{\ell = 1}^N 
\left\vert \partial_{\lambda_\ell} \Psi (\lambda) \right\vert^2 
+ V(\lambda) \left\vert \Psi (\lambda) \right\vert^2 
\right)
\nonumber \\
\left< \Psi \vert \Psi \right> &=& 
\int \prod_{\ell=1}^N \d \lambda_\ell \, \Delta^2 (\lambda)
\left\vert \Psi (\lambda) \right\vert^2,
\nonumber
\eea
the vandermonde determinant $\Delta(\lambda) =
\prod_{\ell < m} (\lambda_\ell - \lambda_m)$
arises as Jacobian of the change of variables, Eq.(\ref{diag}).
The expression suggests to introduce an {\sl antisymmetric} wave function 
$\Phi(\lambda)$:
\bea
\Phi(\lambda) = \Delta (\lambda) \Psi(\lambda_1, \cdots, \lambda_N)
\nonumber
\eea
as the wave function of $N:={\tt dim} {\cal H}$ species of 
first-quantized `analog' {\sl fermions} in one dimensions, 
spanned by the eigenvalues.
The corresponding Schr\"odinger equation is given by
\bea
i {\partial \over \partial t} \Phi(\lambda_1, \cdots, \lambda_N; t)
= \H_{\rm NC} \Phi(\lambda_1, \cdots, \lambda_N; t).
\nonumber
\eea
where the Hamiltonian $\H_{\rm NC}$ is given by: 
\bea
\H_{\rm NC} = \sum_{\ell = 1}^N {\cal H}[\lambda_\ell],
\nonumber
\eea
as a sum of one-particle Hamiltonian ${\cal H}[\lambda]$
\bea
{\cal H}[\lambda]
:= \left[ - {1\over 2 {\rm }\theta} {\partial^2 \over \partial \lambda^2}
+ {\rm }\theta \, V(\lambda) \right].
\label{1ptchamiltonian}
\eea
The Hamiltonian describes non-interacting Fermi gas in an external potential 
$V(\lambda)$.  
The above Hamiltonian is precisely the one derivable from the action 
Eq.(\ref{op3daction}), but in terms of diagonal field variables:
\bea
S = {\rm }\theta \int \d t \sum_{\ell = 1}^N \left[ {1 \over 2}
(\partial_t \lambda_\ell)^2 - V(\lambda_\ell) \right].
\nonumber
\eea
\subsection{Classical, Planar, and Disordered Phases of NCFT$_3$} 
To explore possible disordered phases of the theory, we investigate what sort
of vacuum structure emerges once quantum effects due to the many-body `analog' 
fermions are taken into account. 

For concreteness, consider a potential $V(\lambda)$ with a unique minimum
at $\lambda = 0$, whose classical vacuum is given by 
$\lambda_\ell = 0$ for {\sl all} $\ell = 1, 2, \cdots, N$. 
Harmonic fluctuation around the vacuum is described by the action:
\bea
S_{\rm harm} = {\rm } \theta \int \d t \, 
\sum_{\ell = 1}^N \left[ {1 \over 2} (\partial_t \lambda_\ell)^2 
- \left( V_0 + {1 \over 2}  \Omega^2 \lambda_\ell^2 + \cdots \right) \right],
\nonumber
\eea
where $\Omega^2 := V''(\lambda = 0)$, and the Hamiltonian:
\bea
H_{\rm harm} = \sum_{\ell = 1}^N \left[ {1 \over 2 {\rm } \theta} 
\Pi_\ell^2 + {\rm } \theta \left( V_0 + {1 \over 2} \Omega^2 \lambda^2_\ell \right) \right].
\nonumber
\eea
At classical level, ground-state energy of the vacuum $\lambda = 0$ is given 
by $N {\rm } \theta V_0$ and hence, assuming that $V_0$ is fixed, is of order 
${\cal O}(N {\rm } \theta)$. Quantum mechanically, the ground-state energy 
is increased by the zero-point fluctuations, and is readily estimated by 
applying the Schwarz inequality:
\bea
\left< H \right> &\gsim& N {\rm } \theta V_0 
+ \sum_{\ell = 1}^N 
\left< {\Pi_\ell \over \sqrt{2 {\rm } \theta}}
\cdot \sqrt{{\rm } \theta \over 2} \Omega \lambda_\ell \right>
\nonumber \\
&\sim& N {\rm }\theta V_0 + {1 \over 2} N \Omega.
\eea
The last formula indicates that the quantum effect is of order ${\cal O}(N)$.
One might be content that the result is consistent with what one anticipate 
from the following heuristic argument: for harmonic fluctuation,  
relevant degrees of freedom are the eigenvalues, $\lambda_\ell (t)$. 
As there are $N$ eigenvalues, the zero-point fluctuation is estimated
simply to be $ N \cdot {1 \over 2} \Omega$ and is of order ${\cal O}(N)$. 
If the reasoning is correct, then it implies that, for large noncommutativity
${\rm }\theta \gg 1$, the quantum effects would be completely negligible,
in sharp contrast to $(2+0)$-dimensional case.

It turns out that the above reasoning is incorrect, as Fermi statistics of 
the `analog' fermions are not properly taken into account. We will argue 
momentarily that the quantum effect to the ground-state energy is of order 
${\cal O}(N^2)$ and, based on this, the {\sl quantum} NCFT comprises of three 
distinct phases:
\bea
\underline{\rm classical}, \,\, {\rm GMS~~phase} \,\,: \qquad
{\rm }\theta \, &\sim& \, N^{1 + \nu}
\qquad (\nu > 0) \nonumber \\
\underline{\rm planar}, \, {\rm `t} \,\, {\rm Hooft~~phase} : \qquad
{\rm }\theta \, &\sim& \, 
N \qquad \quad g^2_{\rm eff} = {\rm fixed} \nonumber \\
\underline{\rm disordered} \,\,\,\, \,\,{\rm phase} \,\, \,\,: \qquad
{\rm }\theta \, &\sim& \, N^{1 - \nu} \qquad (\nu > 0).
\label{3limits2}
\eea
To see these phases, it is sufficient to examine the ground-state energy 
at quantum vacua. For simplicity, we will approximate the potential
as a quadratic function with $\Omega = 1$. 
Denoting the one-particle fermion energy levels
as $e_1 \le e_2 \le e_3 \le \cdots$ and the Fermi energy as $e_{\rm F}$, 
the particle number $N$ and the total energy ${\cal E}$ is given by 
\cite{bipz}: 
\bea
N &:=& \sum_{\ell = 1} \theta  \left(e_{\rm F} - e_\ell \right)
\nonumber \\
  &=& \int {\d \lambda \d p \over 2 \pi}
\, \Theta \left( e_{\rm F} - {p^2 \over 2 {\rm } \theta} - {1 \over 2} 
{\rm } \theta \lambda^2 - {\rm }\theta \, V_0 \right)
\nonumber \\
{\cal E} &:=& 
\sum_{\ell = 1} e_\ell \, \Theta \left( e_{\rm F} - e_\ell \right)
\nonumber \\
&=& \int {\d \lambda \d p \over 2 \pi}
\Theta \left( e_{\rm F} - {p^2 \over 2 {\rm } \theta} - {1 \over 2} 
{\rm } \theta  \lambda^2 - {\rm } \theta \, V_0 \right)
\, \left[ {p^2 \over 2 {\rm } \theta} + {1 \over 2} {\rm } \theta 
\lambda^2 + {\rm } \theta \, V_0 \right].
\nonumber
\eea
Here, $V_0$ refers to the minimum of the potential, the {\sl classical} 
energy. 

The above expressions implies that, in the total energy ${\cal E}$,
 the classical contribution is of order ${\cal O}
(N {\rm }\theta)$, while the quantum contribution is of order 
${\cal O} (N^2)$. To show this, solve first the $\Theta$-function
constraint of the `Fermi surface' as 
\bea
\left\vert p (\lambda)  \right\vert \, \le \, \sqrt{2 {\rm } \theta}
\sqrt{ \widetilde{e_{\rm F}} - {1 \over 2} {\rm } \theta \lambda^2}
\qquad {\rm where} \qquad  
\widetilde{e_{\rm F}} := \left(e_{\rm F} - {\rm } \theta \, V_0 \right). 
\nonumber
\eea
It then allows to compute $N$ and ${\cal E}$ explicitly. Begin with the 
particle number, $N$. Integrating over $p$ first, elementary algebra yields 
\bea
N &=& 2 \int {\d \lambda \over 2 \pi}
\sqrt{2 {\rm } \theta} \sqrt{ \widetilde{e_{\rm F}} - {1 \over 2}
{\rm } \theta \lambda^2}
\nonumber \\
&=& \widetilde{e_{\rm F}}.
\nonumber
\eea
This indicates that the Fermi energy 
$\widetilde{e_{\rm F}} = \left(e_{\rm F} - {\rm } \theta V_0
\right)$ is of order ${\cal O}(N)$. We will thus set 
$\widetilde{e_{\rm F}} := N \widetilde{\epsilon}$ and, in the large-$N$
limit, hold $\widetilde{\epsilon}$ fixed to ${\cal O}(1)$ constant. 
Similarly, integrating over $p$ first, the total energy ${\cal E}$ is 
obtained as a sum of classical and quantum contributions:
\bea
{\cal E} = {\cal E}_{\rm classical} + {\cal E}_{\rm quantum},
\nonumber
\eea
where
\bea
{\cal E}_{\rm classical} &=& \widetilde{e_{\rm F}} {\rm }\theta \, V_0
= {\cal O}(N {\rm } \theta )
\label{classicalE} 
\eea
and
\bea
{\cal E}_{\rm quantum}
&=& \Big( {1 \over 4} + {1 \over 4} \Big) \widetilde{e_{\rm F}}^2 
= {1 \over 2} \widetilde{e_{\rm F}}^2
= {\cal O}(N^2).
\label{quantumE}
\eea
The first and the second terms in ${\cal E}_{\rm quantum}$ are 
contributions of kinetic and potential energies, respectively.
Evidently, the result exhibits that ${\cal E}_{\rm quantum}$ is of order 
${\cal O}(N^2)$, not ${\cal O}(N)$ as anticipated from the aforementioned 
naive reasoning. With Fermi statistics taken into account, this correct 
result can be understood intuitively as follows. At $\theta \rightarrow
\infty$ limit, the effect of functional integral measure is to turn the 
eigenvalues into positions of the `analog' {\sl fermions}. As such, because
of the Fermi pressure, the ground-state energy will increase, whose size is 
estimated as  
\bea
\Delta {\cal E} \sim \sum_{\ell = 0}^N {1 \over 2} \ell \Omega 
\sim {\cal O}(N^2),
\nonumber
\eea
thus obtaining the correct scaling in the large-$N$ limit.

>From Eqs.(\ref{classicalE}, \ref{quantumE}), we come to the conclusion that,
in the large-$N$ and large-$\theta$ limit, depending on relative 
magnitude between $N$ and ${\rm }\theta$, the ground-state energy will
scale differently. If $N \gg {\rm }\theta$, the ground-state energy is 
dominated by the classical contribution, which we have referred as the
`classical phase'. If $N \sim {\rm }\theta$, the classical and the quantum
contributions are equally important. This is the `planar phase' -- the phase
familiar in the context of planar expansion of matrix models. If 
$N \ll {\rm }\theta$, the energy is dominated by the quantum contribution,
which we referred to as the `disordered phase'. 

\subsection{Effects of the Gradients}
So far, our analysis was based on truncation of $\Delta \H_{\rm gradient}$ 
term in Eq.\eq{completehamiltonian}. In this section, we will prove that this 
gradient term effect is negligible at weak `t Hooft coupling regime, quite
analogous to the situation for two-dimensional NCFTs analyzed in section 3.   
For the present case, now dealing with temporal evolution, we will proceed
slightly differently and utilize the Gibbs inequality (see, for example,
\cite{sakita}). Begin with the Euclidean partition function, 
expressed in terms of canonically normalized field ${\bf M}(t) = 
\sqrt{\theta} {\bf T}$.
\bea
\Z_N = \int [\d {\bf M}(t) \d {\bf \Pi}(t)]_N \exp \Big( - \int \left[
- i \tr {\bf \Pi}(t) \dot{\bf M}(t) + \H({\bf M}(t)) \right] \d t \Big).
\label{starthere}
\eea
Here, the Hamiltonian $\H$ is given by Eq.\eq{completehamiltonian}, 
which we decompose as
\bea
\H = \H_0 + \Delta \H_{\rm grad}
\nonumber
\eea
where
\bea
\H_0 &=& {1 \over 2}  \tr \Pi^2(t) + {m^2 \over 2} \tr {\bf M}^2.
\nonumber \\
\Delta \H_{\rm grad} &=& {1 \over \theta} \tr \Big( - {1 \over 2}
\left[\widehat{\bf x}_a, {\bf M} \right]^2 \Big).
\label{above1}
\eea
The decomposition allows to estimate the gradient effect nonperturbatively.
To this end, we will apply the Gibbs inequality to the partition function 
Eq.(\ref{starthere}), and obtain 
the following upper-bound to the exact effective action $\Gamma_{\rm exact}$
\bea
\Gamma_{\rm exact} \le \Gamma_0 + \Delta \Gamma .
\nonumber 
\eea
Here,
\bea
\Gamma_0 &=& - \ln \int [\d {\bf M}(t) \d {\bf \Pi}(t)]_N 
\exp \Big( - \int [ - i \tr \Pi (t) 
\dot{\bf M} (t) - \H_0({\bf M}) ]\d t \Big) 
\nonumber \\
\Delta \Gamma &:=& \Big< \int \d t \, \Delta \H_{\rm grad}(t) \Big>_0 
\label{above2}
\eea
where
\bea
\Big< \cdots \Big>_0 = e^{- \Gamma_0} \int 
[\d {\bf M} (t) \d {\bf \Pi}(t)]_N \cdots 
\exp \Big( - i \tr \Pi(t) \dot{\bf M}(t) - \H_0 ({\bf M}) ] \d t \Big).
\nonumber
\eea
The correction $\Delta \Gamma$ is computable utilizing precisely the same 
method as that in section 3.4, except that now the field variables are 
time-dependent \footnote{In fact, for if all the couplings in the 
Hamiltonian are \underline{time-independent}, one can make similarity to the 
method of section 3.4 by utilizing the defining relations:
\bea
\exp(-\Gamma) &=& 
\tr_{{\cal H}_{\rm Fock}} \exp \left( - \beta \, \H \right)
= \tr_{{\cal H}_{\rm Fock}}
\exp \left( - \beta \, \Delta \H_{\rm grad} \right)
\exp \left( - \beta \, \H_0 \right),
\nn
{\rm and} \hskip3cm \exp( - \Gamma_0) &=& \tr_{{\cal H}_{\rm Fock}} 
\exp \left( - \beta \, \H_0 \right).
\nonumber
\eea
Thus, applying the Gibbs inequality, one obtains
\bea
\Gamma_{\rm exact} \le \Gamma_0 + \beta \Big< \Delta \H_{\rm grad} \Big>_0 ,
\nonumber
\eea
where
\bea
\Big< (\cdots) \Big>_0
:= { \tr_{{\cal H}_{\rm Fock}} (\cdots) \exp \left( - \beta , \H \right)
\over \tr_{{\cal H}_{\rm Fock}} \exp \left( - \beta \, \H \right) }
.
\label{alternative}
\eea
Evaluation of Eq.\eq{alternative} is achievable precisely as in the
two-dimensional
Euclidean NCFTs, as the latter can be viewed as the \underline{classical}
statistical mechanics of the $(2+1)$-dimensional NCFTs. 
Thus, utilizing the results of section 3.4, we obtain the same
results and conclusions as in  Eq.\eq{thisisit}.} .
This renders the two-point propagator 
$\Big< {\bf M}_{kl}(t) {\bf M}_{mn}(t') \Big>$ behaves
for short time differences $| t - t'|$, as $\sim
\exp(- m \vert t - t' \vert)$. Fortuitously, computation of $\Delta \Gamma$
involves only {\sl coincident} two-point propagator 
(see Eqs.\eq{above1},\eq{above2}), and involves precisely the
same group theoretic combinatorics as in Eq.\eq{correlator}.
Thus, following the same large-$N$ counting as in section 3.4, we obtain
\bea
\Delta \Gamma &=& \Big<{1 \over \theta} 
\tr \Big( - {1 \over 2} \left[\widehat{\bf x}_a, {\bf M} \right]^2 \Big)
\Big>  
\nonumber \\
&\sim& {N^2 \over \overline{\theta}}  \left( 
\int d\lambda\ \rho(\lambda)\ \lambda^2 - 
\Big[\int d\lambda\ \rho(\lambda)\ \lambda\Big]^2 \right)
\nonumber \\
&\sim& {1 \over \overline{\theta}} {\cal O}(N^2).
\label{thisisit}
\eea
We conclude that, nonperturbatively, size of the gradient effect is 
bounded from above, and is suppressed by $1/\overline{\theta}$ compared 
to the estimates based on Hermitian matrix quantum mechancs with Hamiltonian $\H_0$.  

\subsection{Quantum Vacua and Solitons}
Having identified the three possible phases at quantum level, we now
examine vacua and solitons, and their quantum aspects.
Introduce the second-quantized fermion field, $\Psi(x, t)$. 
The Hamiltonian is then expressible as
\bea 
H = \int \d \lambda \, \Psi^\dagger (\lambda, t)
\left( - {1 \over {\rm }\theta} {\partial \over \partial \lambda^2}
+ {\rm } \theta \, V(\lambda) \right) \Psi(\lambda, t),
\label{2ndhamiltonian}
\eea
and interpret it as the Hamiltonian for a second-quantized fermion interacting
with the external potential, $V(\lambda)$.  
In the saddle-point approximation, the equation of motion of 
the density field $\rho(\lambda, t)$ is given by\footnote{This equation
of motion is approximate \cite{dmw}, though it is sufficient
for our present purpose.}
\bea
\partial_t \left({1 \over \rho} \partial_t \partial^{-1}_\lambda
\rho \right) = \partial_\lambda \left({1 \over 2} \rho^2 + V(\lambda) \right).
\label{integral}
\eea
Utilizing the WKB approximation for the energy levels, one finds the 
static solution of the density field as:
\bea 
\rho_{\rm s} (\lambda) &=& \left\{
\begin{array}{ccc}
N^{_1} g_{\rm eff}^{-2} \sqrt{2\left(g^2_{\rm eff} E -V(\lambda) \right)}&
\qquad \rm{for} \qquad V\left(\lambda \right) \, \le \, g^2_{\rm eff} E
\label{saddlepoint}
\\ &\\ 0 & \rm{otherwise} & \\ \end{array} \right. ,
\label{rhodistrn}
\eea
where $g^2_{\rm eff}$ refers to the `t Hooft coupling parameter,
$g^2_{\rm eff} = (N/ {\rm }\theta)$ and $E$ refers to the first-integral 
of Eq.(\ref{integral}), piecewise constant on each classically allowed
region and is fixed by the normalization condition:
\bea
\int\limits_{-\infty}^{+\infty} \d \lambda \, \rho_{\rm s}
(\lambda) \,= 1 .  
\label{condition}
\eea
Thus, in the case of double-well potential, taking the first integrals 
of motion, $E_1, E_2$, on the left and the right wells, respectively, to be 
below the energy at the top of the potential, the static density field 
$\rho_{\rm s} (\lambda)$ is supported at the two disconnected parts 
(see Figure 5) -- ${\cal D}_{\rm L}, {\cal D}_{\rm R}$, respectively. 
The normalization condition Eq.(\ref{condition}) then implies that
\bea
\int_{{\cal D}_{\rm L}}\d \lambda\, \rho_{\rm s} (\lambda) = n_1 
\qquad {\rm and} \qquad
\int_{{\cal D}_{\rm R}}\d \lambda\, \rho_{\rm s} (\lambda) = n_2
\qquad {\rm where} \qquad n_1 + n_2 = 1.
\nonumber
\eea
The two extreme limits, $n_1 = 0$ and $n_2 = 0$, correspond to the two 
`quantum' vacua, distributed around the respective locations of the 
classical vacua, while nonzero pairs of $[n_1, n_2]$ correspond to the 
`quantum' solitons. 
Note that the first-integrals of motion, $E_1, E_2$, take different values
generically, as quantum-mechanical tunnelling between the two potential wells 
is suppressed in the $N, {\rm }\theta \rightarrow \infty$ limit.
\begin{figure}[htb]
\begin{center}
\epsfxsize=15cm\leavevmode\epsfbox{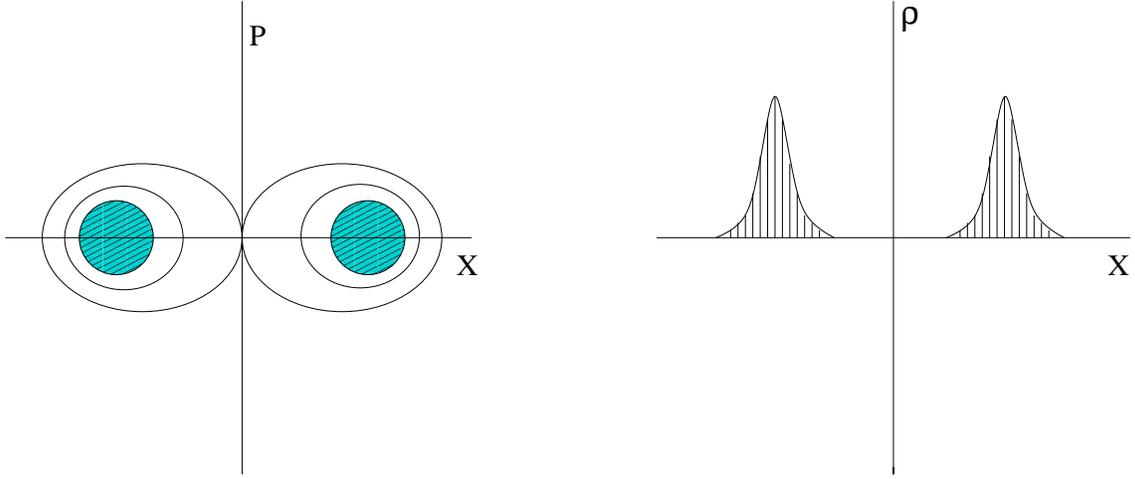}
\end{center}
\caption{\sl Density profile of the `analog' fermions in the one-particle
phase-space and in the eigenvalue space. The width of density profile is 
given by $\Delta \lambda \sim 
\left(\widetilde{e_{\rm F}}/\theta \right)^{1/2}$ 
and $\Delta p \sim 
\left(\widetilde{e_{\rm F}} \theta \right)^{1/2}$ so that 
$\Delta \lambda \Delta p \sim \widetilde{e_{\rm F}}$, consistent with 
Fermi statistics. In the classical limit, the profile reduces to 
delta-function distributions.}
\label{fermi-fig}
\end{figure}

\subsubsection*{Classical Limit: $\hbar \rightarrow 0$}
As a consistency check of the aforementioned three phases of the {\sl quantum}
NCFT, we will now examine the {\sl classical} limit by taking $\hbar \to 0$, 
while holding $N$ large but fixed. First, from Eq.(\ref{3limits2}), we 
observe that the Planck constant $\hbar$ ought to be associated with 
$g^2_{\rm eff}$, as taking $g^2_{\rm eff} \equiv (N/{\rm } \theta)
\rightarrow 0$ along with $N \rightarrow \infty$ renders the planar-phase
to approach to the classical phase. In the notation of Eq.(\ref{3limits2}),
this implies that the `t Hooft coupling scales as $g^2_{\rm eff} \sim 
N^{- \nu} \rightarrow 0$. Hence, in this subsection, we will take the Planck 
constant $\hbar$ equal synonymously to the `t Hooft coupling $g^2_{\rm eff}$.

Consider the double-well potential studied in the previous subsections. 
In the classical limit, we expect that profiles of the 
eigenvalue density field is reduced to those of classical phase, viz. 
vacua and solitons found in \cite{gms}. This can be understood as follows. 
For $E_1, E_2$ below the potential barrier, each disconneced support of the
eigenvalue density field $\rho_{\rm s}(\lambda) $ in Eq.(\ref{saddlepoint}) 
shrinks 
as $\hbar \sim g^2_{\rm eff} \rightarrow 0$, to a certain distribution of 
zero width, centered around $\lambda=\lambda_1$ and $\lambda=\lambda _2$, 
respectively. Examining the limit carefully, we find that
\bea
\frac{\rho_{\rm s} (\lambda)}{\hbar } \quad \longrightarrow \quad
n_1 \delta (\lambda-\lambda _1) +n_2 \delta(\lambda-\lambda _2) ,
\nonumber
\eea
reproducing accurately the classical profile of the eigenvalue density field, 
Eq.(\ref{classicaldistrn}).
Stated differently, starting from classical vacua and solitons 
Eq.(\ref{classicaldistrn}) of Gopakumar,
Minwalla and Strominger \cite{gms}, turning on the quantum effects renders
them into filled Fermi sea of the Hermitian matrix quantum mechanics, either
on a single well or multiple wells. See fig \ref{fermi-fig}.

To convince the readers that the classical limit is reproducible correctly, 
consider the simplest situation again -- the single well potential
$V (\lambda) =\frac{1}{2} \Omega^2 \lambda^2$. Then, at large-$N$ limit, 
$E= N \hbar \Omega$ (ignoring the zero-point fluctuation energy), and 
Eq.(\ref{rhodistrn}) implies that 
\bea
\int\limits_{-\infty}^{+\infty} \d \lambda \, \rho_{\rm s} (\lambda) \, = \, 
\frac{1}{N \hbar} \int\limits_{-\lambda_0}^{+\lambda_0} \d \lambda 
\sqrt{2\left(N \hbar \Omega - \frac{1}{2} \Omega^2 \lambda^2 \right)} = 1
\nonumber
\eea
over the band $[-\lambda_0, + \lambda_0]$. Here, we have used the following 
value of the turning point, defining the band-edge of the distribution: 
\bea
\lambda_0= \sqrt{2 N \hbar / \Omega}.
\nonumber
\eea
One immediately notes that Planck's constant $\hbar$ is in the right place
in Eq.(\ref{rhodistrn}), identifiable as $g^2_{\rm eff} \sim \hbar$.
As $\hbar \rightarrow 0$, for a fixed but large $N$, the band-edge $\lambda_0
$ scales to zero, and hence the band width shrinks to zero size.
At the same time, the mid-band density scales as $\sqrt{\Omega / N \hbar} 
\rightarrow \infty$. Evidently, in the classical limit, product of the 
mid-band density times the band width remains constant and is always of order 
${\cal O}(1)$.

As is well exploited in the context of matrix model description of $c=1$
noncritical string \cite{c=1}, 
profile of the density field $\rho(\lambda)$ is expressible alternatively using Wigner's phase-space distribution function of the $N$ 
`analog' fermions: 
\bea
{\rm F}(p,\lambda;t)\; &=&\,\int \d x \,
\Psi^{\dagger }\left(\lambda-{\hbar \over2}x, t \right)\,
e^{{i \over \hbar} px }\,\Psi \left(\lambda+{\hbar \over 2} x, t \right)
\nonumber \\
&=&\,\int \d x \,\,\Psi^{\dagger }\left(\lambda,t\right) \,\star
\, e^{{i \over \hbar} px }\star \Psi\left(\lambda,t\right).
\nonumber
\eea
Here, the coordinates $(\lambda, p)$ obey Moyal's commutation relation,
$[p,\lambda]_{\star }=i\hbar \,\,$%
\footnote{It is worthy noting that the matrix model for $c=1$ noncritical
string provides an early example of noncommutative field theory.}.  
In terms of Wigner's distribution function,
the eigenvalue density field is expressed compactly as:
\bea
\rho (\lambda,t) = {\hbar \over N} \int \d p\, {\rm F}(p,\lambda;t),
\nonumber
\eea
measuring the distribution of the eigenvalues. The factor of $\hbar$ 
reproduces correctly the normalization condition 
$\int \d p \d \lambda \, \hbar {\rm F}(p,\lambda; t) = 1$.
As shown in \cite{dmw}, 
the Wigner's function corresponding to saddle-point 
configuration is simply given in the first-quantized description by the 
phase-space density of $N$ fermions. 
The fermions occupy the lowest $N$ energy 
eigenstates of the one-particle Hamiltonian ${\cal H}(\lambda)$ in 
Eq.\eq{1ptchamiltonian}. 

\subsection{Second-Quantized Description}
Actually, using the second-quantized fermion field operators introduced in
Eq.(\ref{2ndhamiltonian}), the eigenvalue density field is now expressible as
the fermion number density operator:
\bea
\hat\rho(\lambda, t) = {1 \over N}
 \Psi ^{\dagger }(\lambda,t)\,\Psi \left(\lambda,t\right),
\label{temp1}
\eea
yielding correct normalization 
$ \int \d \lambda \, \Psi^\dagger(\lambda, t) \Psi(\lambda, t) = N$.
Taking expectation value of Eq.(\ref{temp1}) on a many-particle state 
$\vert \lambda_1, \ldots, \lambda_N \rangle$, antisymmetrized product
of $N$ position eigenfunctions, we obtain 
\begin{equation}
\left< \lambda_1, \ldots, \lambda_N \vert \, \hat \rho(\lambda) \,
\vert \lambda_1, \ldots, \lambda_N \right> = {1 \over N} \sum_{\ell=1}^{N} 
\delta \Big( \lambda - \lambda_\ell (t) \Big),
\label{rho-vev}
\end{equation}
matching perfectly with Eq.(\ref{rho-def}). It also satisfies the normalization
condition 
\begin{eqnarray}
\int\limits_{-\infty}^{+\infty} \d \lambda \, \left< \, 
\hat \rho(\lambda, t) \, \right> = 1.
\nonumber
\end{eqnarray}
The Eq.(\ref{rho-vev}) implies that $\hat \rho (\lambda)$ operator is
expressible as
\begin{equation}
\hat \rho(\lambda) = {1 \over N} \sum_{\ell=1}^N 
\delta \Big(\hat \lambda - \lambda_\ell (t) \Big),  
\label{temp2}
\end{equation}
where $\hat \lambda$ refers to the position operator in the 
first-quantized description. This then defines the density field operator 
at quantum level.

Equipped with the eigenvalue density field operator Eq.(\ref{temp2}) 
via the second-quantized fermion field $\Psi$, one can exploit
quantum effect to the NCFT vacua and solitons.
Restricting low-energy excitations to the U$(\infty)$ invariant sector, we have
found that classical dynamics of the tachyon field is described by the density 
field $\rho(\lambda)$. Likewise, in the same U$(\infty)$ invariant sector, 
quantum dynamics of the tachyon field is described by the density field 
operator $\hat(\hat \lambda)$ defined in Eq.(\ref{temp2}).
The extent of quantum effects can be judged by taking expectation value of 
Eq.(\ref{temp2}) and measuring deviation from its classical value
Eq.(\ref{rho-def}). For instance, by approximating the eigenvalue density 
field operator to be the same as the classical distribution, we have obtained
in the previous subsection that 
\begin{eqnarray}
\Big< \hat \rho(\lambda) \Big> = \frac{1}{\hbar} \sqrt{ 2(E - V(\lambda))}.
\nonumber
\end{eqnarray}

Equivalently, U$(\infty)$ invariant information of the tachyon field is 
governed by the change of variable:
\bea
{1 \over N} \tr_{\cal H} {\bf T}^n = \int\limits_{-\infty}^{+\infty} \d \lambda\,\lambda^n \rho(\lambda).
\nonumber
\eea
Thus, from a knowledge of the classical $\rho(\lambda)$, one can reconstruct 
the classical tachyon field $\bf T$ in the Weyl formulation. One can 
subsequently rebuild the tachyon field $\T (x)$ on $\real^2$ via the 
Weyl-Moyal correspondence map. The reconstruction is equally applicable
at quantum level. For instance, 
\bea
{1 \over N} \tr_{\cal H} 
\left< \widehat{\bf T}^n \right> = \int\limits_{-\infty}^{+\infty} 
\d \lambda \, \lambda^n \Big< \widehat{\rho}(\lambda) \Big>.
\nonumber
\eea
Denote the image of the Weyl-Moyal map of $\Big< \widehat{\bf T} \Big>$ 
as $\widehat{ \T} ({\bf x})$. Then, the above equation becomes 
\bea
{1 \over V(\real^2_\theta)} \int_{\real^2_\theta} d^2 {\bf x} \, \left[ 
\widehat{\T}({\bf x}) \star \widehat{\T}({\bf x}) \star
\cdots \star \widehat{\T}({\bf x}) \right]_{\rm n-tuple} = 
\int\limits_{-\infty}^{+\infty} \d \lambda \lambda^n 
\Big< \widehat{\rho}(\lambda) \Big>.
\label{maps}
\eea
This moment relation enables to reconstruct the `quantum' profile of
the vacua and solitons over $\real^2_\theta$ We will draw utility of the
map by illustrating two representative physical consequences driven by
the `quantum effects'. 
\subsubsection*{\hskip1cm Quantum Destruction of Long-Range Order}
We have already demonstrated that the quantum effect drives the classical
density profile of delta-function type into a Fermi distribution, as 
depicted in Fig.\eq{fermi-fig}. A consequence of broadening into 
the Fermi distribution is that the translational invariance over 
$\real_\theta^2$, viz. ${\bf x}^a \rightarrow {\bf x}^a + $ (constant), 
which is respected by all classical vacua, is dynamically broken. 

Recall that the classical vacua correspond to density distribution of 
delta-function type, all eigenvalues taking the same value, say, $\T_0$. 
Thus, Eq.(\ref{maps}) yields
\bea
{1 \over V(\real^2_\theta)} \int_{\real^2_\theta} d^2 {\bf x} \,
\Big( \widehat{\T}({\bf x}) \Big)^n_\star = \T_0^n
\qquad {\rm for} \qquad n = 1, 2, 3, \cdots.
\nonumber
\eea
and hence find the unique solution as 
$\widehat{\T}({\bf x}) = \T_0$ --- a {\sl homogeneous} configuration,
respecting the translational invariance over $\real^2_\theta$. 

Once quantum effects are taken into account, as shown above, the classical
delta-function type density distribution is broadened into a 
Fermi distribution putting each eigenvalue at a different value 
from one another -- a consequence of repulsion between adjacent eigenvalues. 
In this case, it is fairly straightforward to convince oneself that there is
{\sl no} homogeneous solution solving the moment map Eq.(\ref{maps}) for
{\sl all} $n$. As such, a generic solution of Eq.(\ref{maps}) ought to be a 
nontrivial function over $\real^2_\theta$. To illustrate this, let us take 
\bea
\Big< \widehat{\rho}(\lambda) \Big> = \left\{ \begin{array}{ccc}
1/R & {\rm for} & -R/2 \le \lambda \le +R/2 \\
0 & {\rm otherwise} & \\
\end{array} \right.  .
\nonumber
\eea
Then, a solution of Eq.(\ref{maps}) is easily found as
\bea
\widehat{\T}({\bf x}) = \left\{ \begin{array}{ccc}
x^1 & {\rm for} & -R/2 \le x^1 \le +R/2 \\
0 & {\rm otherwise} & \end{array} \right. ,
\label{solution}
\eea
thus breaking translational invariance along the $x^1$-direction over 
$\real^2_\theta$, though invariant under translation along the
$x^2$-direction. There are also infinitely many other solutions to 
Eq.(\ref{maps}), including the `stripe-phase' states, 
but they are all related to the solution 
Eq.(\ref{solution}) via U($\infty$) rotations, viz. solutions of the type 
$U({\bf x}) \star \T(x^1) \star U^{-1}({\bf x})$ 
for an arbitrary ${\rm U}({\bf x})$.

We conclude that the translational long-range order of the classical vacua 
in NCFT is destroyed generically by quantum fluctuations. 

\subsubsection*{\hskip1cm Quantum Corrected Soliton Mass}
Consider the classical soliton of type $(N_a, N_b)$, Eq.\eq{soliton}. We
are interested in estimating the mass of the `quantum' soliton, or,
equivalently, the quantum correction to the soliton mass. Take, for
definiteness, the potential $V({\bf T})$ of the type given in 
Fig.\eq{regularization}.  Classically, the soliton mass is simply given by 
the increase of the potential energy by moving, out of total $N = N_L +N_L$
eigenvalues situated at the global vacuum on the left well, 
a fraction of $N_R$ eigenvalues to the local minima on the right well.
Denoting the energy difference between the left and the right
wells, $V_L, V_R$,  as $\Delta V = (V_R - V_L)$, the classical soliton mass 
is given by
\bea
M_{(N_L, N_R)}[{\rm classical}] \quad \sim \quad \theta N_R \Delta V.
\nonumber
\eea
Quantum mechanically, eigenvalue distribution for both the global vacuum
and the $(N_L, N_R)$ soliton will be broadened into Fermi distributions. 
Thus, the quantum corrected soliton mass is estimated by computing the
difference of the energy functional averaged over the Fermi distributions
according to Eq.(\ref{maps}). Utilizing the results of 
Eqs.(\ref{classicalE}, \ref{quantumE}), 
we estimate the quantum corrected soliton mass as
\bea
M_{(N_L, N_R)}[{\rm quantum}] 
&=& \Big< {\cal E}_{(N_L, N_R)} \Big> - \Big< {\cal E}_{(N_L + N_R, 0)} \Big>
\nonumber \\
&\sim& \left( {\rm } \theta N_L V_L + N_L^2 \right)
+ \left( {\rm } \theta N_R V_R + N_R^2 \right)
- \left( {\rm } \theta N V_L + N^2 \right)
\nonumber \\
&=&  M[{\rm classical}] - 2 N_L N_R.
\nonumber
\eea
We thus deduce that the quantum correction, as given by the second term in 
the last expression, is {\sl negative} and is of order ${\cal O}(N^2)$.
Evidently, the correction is negligible in the GMS-phase, comparable in
the planar-phase, but outweighs the classical mass in the disordered phase.
\section{Discussions}
Before closing, we would like to bring up investigation of related
phenomena in other contexts. The first is concerning quantum effects
either in IKKT Type IIB or in BFSS Type IIA matrix theories. For Type
IIB IKKT matrix model, the issue of measure-induced interaction
between eigenvalues and its consequences have been considered
previously, albeit in different context and with different motivation.
See, for instance, results of \cite{Ambjorn:2001xs} and references
therein.  The classical moduli space is given by ten commuting
matrices whose eigenvalues span $\real^{10}$, ten-dimensional
Euclidean spacetime.  A calculation of the matrix partition function
indicates that the moduli space is partly lifted and, morally
speaking, a smaller-dimensional submanifold remains nocompact and
flat. The result is attributed to a logarithmic interaction between
eigenvalues as the remaining ``angular'' degrees of freedom are
integrated out.  This is similar to the vandermonde effect of the
one-matrix model.

Classical solutions of the IKKT and BFSS matrix models include all of
the D-branes in Type IIA and IIB strings. The low-energy theory is
equivalent to NCFTs involving both scalar and gauge fields.  An
immediate question is whether there exist various kinds of large-$N$
limits in these field theories, some of which might destabilize the
D-branes by quantum fluctuations.  We believe that this is a very
important issue, so let us remark a little further. One place to look
for this sort of effect would be one-loop computations in the IKKT and
BFSS matrix models which might show the necessity of a sort of
'tHooft-like scaling, viz. $\theta \sim N$, without which the D-brane
solutions might be completely destabilized, which is the counterpart
of the disordered phase studied in this paper. Of course, for D$p$
branes or a system of D$p$-D$q$ branes (with $p=q$ mod 4) bosonic and
fermionic determinants cancel because of supersymmetry and there are
no large-$N$ divergences at 1-loop. On the other hand, supersymmetry
is broken in situations involving (i) relative motion between the BPS
branes, (ii) D$p$-D$q$ branes with $p-q$ not a multiple of 4, and
(iii) brane-antibrane systems. In (iii), the $D2-\bar D2$ system
\cite{braneantibrane,ddbar} was studied extensively, and it would be
an interesting starting place to address the large-$N$ issues raised
here.

Second, as elaborated in section 2.7, the measure effect we have
discussed in this paper is intimately related to the phenomenon of IR
divergence \cite{uvir} through nonplanar diagrams. Recently, it has been 
shown \cite{reys} that the completion of all the nonplanar diagrams
participating in the UV-IR mixing in NCFTs studied in this work is
expressible entirely in terms of scalar counterpart of the open Wilson
lines \cite{owl}. The effective action then interpreted as (Legendre
transform of) an effective field theory of noncommutative dipoles --
noncommutative manifestation of dynamically generated `closed strings'
\cite{rey}.  There, the result was based exclusively on the Moyal
formulation. An interesting problem is to recast the result in Weyl
formulation, and to understand the three different scaling regimes in
terms of the open Wilson lines and noncommutative dipoles.

Finally, it would be interesting to see if the transition to the
disordered phase discussed in this paper is related to the large $N$
phase transition \cite{gww}.

We will report progress regarding the above  problems elsewhere.

\subsection*{Acknowledgement}
We are grateful to S.R. Das, A. Dhar, M.R. Doulgas, R. Gopakumar,
D.J. Gross, V. Kazakov, S. Minwalla, S. Mukhi, A. Sen, and
S.H. Shenker for enlightening discussions.  We would like to thank
hospitality of Theory Division at CERN (GM and SJR), 
Institut Henri Poincar\'e (SJR), and Institut des Hautes \`Etudes 
Scientifiques (SJR) during this work. 

\appendix
\section{Weyl-Moyal correspondence}
In this section we briefly review the operator formulation of NCFT 
in the context of Sec 2.1.

One begins by introduce an `auxiliary' one-particle Hilbert space 
${\cal H}$, of dimension ${\tt dim}{\cal H}= N$, \footnote{%
Since representations of Eq.\eq{heisen} are necessarily infinite-dimensional,
$N=\infty$ at the moment. We will shortly discuss
(Sec  2.3) how on a
noncommutative torus with rational $\theta$, $N$ becomes finite.}%
carrying a representation of the Heisenberg algebra:
\bea
\left[ \widehat{x^a}, \widehat{x^b} \right] = 
i \epsilon^{ab} \identity .
\nonumber
\label{heisen}
\eea
The Weyl-Moyal map refers to isomorphism between
functions on $\real^2_\theta$ and operators on ${\cal H}$:
\begin{eqnarray}
{\bf x} \quad &\longleftrightarrow &\quad \widehat{\bf x} \nonumber\\
\T({\bf x})\quad &\longleftrightarrow &\quad {\bf T} (\widehat{{\bf x}}) 
\nonumber \\
V_{\star}(T)\quad &\longleftrightarrow &\quad V({\bf T}) 
\nonumber \\
\int \d^2{\bf x}\cdots \,\, &\longleftrightarrow &\quad {\tr}_{{\cal H}%
_\theta }\cdots ,  \label{wm-map}.
\end{eqnarray}
In particular, in plane-wave basis, the Weyl-Moyal map renders the following
one-to-one correspondence between fields: 
\bea
\T({\bf x}) = \int_{\widetilde{\hskip0.2cm \real^2}} 
{\d^2 {\bf k} \over (2 \pi)^2} \, e^{ - i {\bf k} \cdot {\bf x}} \, 
\tr_{{\cal H}} \left( e^{ i {\bf k}\cdot \widehat{\bf x}} \, 
{\bf T} \right),
\eea
which follows from Weyl-ordering prescription of the operators 
$\widehat{\bf x}$'s.

The map, Eq.(\ref{wm-map}), then equates the NCFT action 
Eq.(\ref{expand2daction}) with Eq.\eq{weylaction}.
Operators on ${\cal H}$ are realizable in terms of matrices once
we introduce a complete set of orthonormal basis of ${\cal H}$ as 
$\vert \ell \rangle$, $\ell = 1, 2, \cdots, {\tt dim}{\cal H}
\equiv N$, and
one-dimensional projection operators therein: 
\begin{eqnarray}
{\bf P}_\ell \,\, = \,\, \vert \ell \rangle \langle \ell \vert \qquad 
\qquad \ell = 1, 2, \cdots, {\tt dim}{\cal H}
\equiv N.  \label{projop}
\end{eqnarray}
The ${\bf P}_\ell$'s satisfy the projective and the completeness relations: 
\begin{eqnarray}
{\bf P}_\ell {\bf P}_m \,\, = \,\, \delta_{\ell m}{\bf P}_m \qquad \quad 
{\rm and} \quad \qquad \sum_{\ell =1}^{N} 
{\bf P}_\ell = \identity.  \nonumber
\end{eqnarray}

At leading order in $(1/{\rm }\theta)$, both Eq.(\ref{expand2daction}) and
(\ref{weylaction}) are invariant under 
\bea
\T({\bf x}) \rightarrow U({\bf x}) \star \T({\bf x}) \star U^{-1}({\bf x})
\qquad \longleftrightarrow \qquad
{\bf T} \rightarrow {\bf U} {\bf T} {\bf U}^{-1},
\label{uinfinity}
\eea
representing area-preserving diffeomorphism, equivalently, U$(\infty)$
symmetry. The symmetry is broken explicitly by the term ${\cal L}_{-1}$.
\section{Large $N$ Saddle-Point of One-Matrix Model}  
As mentioned in Sec 2.4, taking $\theta= N/g^2_{\rm eff}$
and small enough $g_{\rm eff}$, we have seen that 
the large-$N$  saddle-point for the density  $\rho_s$  
($\rho$ defined in Eq.\eq{density}) is simply an extremum of the
effective action Eq.\eq{collaction} (with 
the constraint Eq.(\ref{constraint}) taken care
of by a lagrange multiplier $E$):
\bea
S_{\rm total}[\rho] = S_{\rm eff} [\rho] + E 
\Big( 1 - \int_{\cal D} \d \lambda  \rho(\lambda) \Big).
\label{stotal}
\eea
The saddle-point equation for $\rho$ then reads:
\bea
\partial_\rho S_{\rm total}[\rho]
= N^2 \left[ g^{-2}_{\rm eff}
V(\lambda) - 2 \int_{\cal D} \d \mu \rho(\mu) \ln \vert \lambda
- \mu \vert \right] - N \theta E = 0 \qquad
{\rm for} \quad \lambda \subset {\cal D},
\label{saddleeqn}
\eea
viz. analytically continuing to complex-$\lambda$ plane, the real-part of
\bea
V_{\rm eff} (\lambda) = V(\lambda) - 2 g^2_{\rm eff} \int_{\cal D} \d \mu
\, \rho(\mu) \ln (\lambda - \mu)
\nonumber
\eea
remains constant $E$ over the support ${\cal D}$. Taking 
derivative of Eq.(\ref{saddleeqn}) with respect to $\lambda$, 
one obtains the following dispersion-relation:
\bea
{1 \over 2 g^2_{\rm eff}} V'(\lambda) = 
\int_{\cal D} \hskip-0.5cm - \hskip0.4cm 
\d \mu \, {\rho(\mu) \over \lambda - \mu} \qquad
{\rm for} \quad \lambda \subset {\cal D}.
\nonumber
\eea
The right-hand side is related to the resolvent 
${\cal R}(\lambda)$ of the eigenvalue distribution:
\bea
{\cal R}(\lambda) := \lim_{N \rightarrow \infty}
\left< {1 \over N} \tr {1 \over \lambda - {\bf T}} \right>
= \int_{\cal D} \d \mu \, {\rho (\mu) \over (\lambda - \mu)},
\qquad
{\rm Re} {\cal R}(\lambda) = \int_{\cal D} \hskip-0.5cm - \hskip0.4cm
\d \mu \, {\rho(\mu) \over (\lambda - \mu)},
\label{kernel}
\eea
supplemented with the boundary condition
\bea
{\cal R}(\lambda) = {1 \over \lambda} + {\cal O}\left({1 \over \lambda^2} 
\right) \qquad {\rm for} \qquad \lambda \rightarrow \infty 
\nonumber
\eea
as the consequence of the normalization condition 
\bea
\int \d \lambda \, \rho(\lambda) = 1.
\nonumber
\eea

Consider now the potential Eq.\eq{doublewell}. Let us
look for a saddle-point corresponding to the 
$(N_1, N_2)$-instantons. Evidently, extending the above
results, quantum counterpart of $(N_1, N_2)$
instantons ought to correspond to so-called two-cut distributions in matrix 
model. The two-cut distribution is characterized by two disjoint intervals
${\cal D}_1, {\cal D}_2$ and fractions of eigenvalue density:
\bea
n_1 = \int_{{\cal D}_1} \d \lambda \, \rho(\lambda)
\quad {\rm and} \quad
n_2 = \int_{{\cal D}_2} \d \lambda \, \rho(\lambda)
\qquad  {\rm with} \quad n_1 + n_2 = 1.
\nonumber
\eea
At large-$N$, large-$\theta$ limit, the total action is now given by
\bea
S_{\rm total} [\rho; n_1, n_2]
= S_{\rm eff}[\rho]
+ E_1 \left(n_1 - \int_{{\cal D}_1} \d \lambda \, \rho(\lambda) \right)
+ E_2 \left(n_2 -\int_{{\cal D}_2} \d \lambda \, \rho(\lambda) \right).
\nonumber
\eea
The saddle-point equation for $\rho(\lambda)$ takes the same form as before,
viz.:
\bea
N^2 \left( g^{-2}_{\rm eff} V(\lambda) - 2  
\int_{{\cal D}_1} \d \mu \rho(\mu) \ln \vert \lambda - \mu 
\vert \right)
&=& N \theta E_1 \qquad \quad {\rm for} \qquad \lambda \subset {\cal D}_1
\nonumber \\
N^2 \left( g^{-2}_{\rm eff} V(\lambda) - 2 
\int_{{\cal D}_2} \d \mu \rho(\mu) \ln \vert \lambda - \mu 
\vert \right)
&=& N \theta E_2 \qquad \quad {\rm for} \qquad \lambda \subset {\cal D}_2.
\nonumber 
\eea
The saddle-point equation with respect to $n_1$ yields:
\bea
\partial_{n_1} S_{\rm total}[\rho; n_1, n_2]
= \Big( E_1 - E_2 \Big) = 0.
\label{saddleeqn2}
\eea
For the double-well potential of the type Eq.(%
\ref{doublewell}), the distribution on the two wells is symmetric. Using the 
methods mentioned above (see \cite{bipz, bmn}
for more details), one
finds that the saddle-point is given in terms of two-cut eigenvalue
distribution Eq.\eq{distrn}.


\begin{thebibliography}{9}

\bibitem{cds}
A.~Connes, M.~R.~Douglas and A.~Schwarz,
JHEP {\bf 9802}, 003 (1998)
[hep-th/9711162]/

\bibitem{sw}
N.~Seiberg and E.~Witten,
JHEP {\bf 9909}, 032 (1999)
[hep-th/9908142].

\bibitem{gms}  
R.~Gopakumar, S.~Minwalla and A.~Strominger,
JHEP {\bf 0005}, 020 (2000)
[hep-th/0003160].

\bibitem{gms2}
M.~Aganagic, R.~Gopakumar, S.~Minwalla and A.~Strominger,
JHEP {\bf 0104}, 001 (2001)
[hep-th/0009142].

\bibitem{rocek}
L.~Hadasz, U.~Lindstrom, M.~Rocek and R.~von Unge,
JHEP {\bf 0106}, 040 (2001)
[hep-th/0104017].

\bibitem{gopa2}
R.~Gopakumar, M.~Headrick and M.~Spradlin,
[hep-th/0103256].

\bibitem{ns}
N.~Nekrasov and A.~Schwarz,
Commun.\ Math.\ Phys.\  {\bf 198}, 689 (1998)
[hep-th/9802068].

\bibitem{gn}
D.~J.~Gross and N.~A.~Nekrasov,
JHEP {\bf 0007}, 034 (2000)
[hep-th/0005204].

\bibitem{poly}
A.~P.~Polychronakos,
Phys.\ Lett.\ B {\bf 495}, 407 (2000)
[hep-th/0007043].

\bibitem{jatkar}
D.~P.~Jatkar, G.~Mandal and S.~R.~Wadia,
JHEP {\bf 0009}, 018 (2000)
[hep-th/0007078].

\bibitem{mukhi}
K.~Dasgupta, S.~Mukhi and G.~Rajesh,
JHEP {\bf 0006}, 022 (2000)
[hep-th/0005006].

\bibitem{wittentachyon}
E.~Witten,
[hep-th/0006071].

\bibitem{harveyetal}
J.~A.~Harvey, P.~Kraus, F.~Larsen and E.~J.~Martinec,
JHEP {\bf 0007}, 042 (2000)
[hep-th/0005031].

\bibitem{mandalrey}
G.~Mandal and S.-J.~Rey,
Phys.\ Lett.\ B {\bf 495}, 193 (2000)
[hep-th/0008214].

\bibitem{bipz}
E.~Brezin, C.~Itzykson, G.~Parisi and J.~B.~Zuber,
Commun.\ Math.\ Phys.\  {\bf 59}, 35 (1978).

\bibitem{matrixmodel}
E.~.~Brezin and S.~R.~Wadia,
{\sl The Large N expansion in quantum field theory and statistical
physics: From spin systems to two-dimensional gravity} (World Scientific,
Singapore, Singapore, 1993);\\
M.L. Mehta, {\sl Random matrices}, 2nd ed. (Academic Press, New York, 1991);\\
P.A. Mello, {\sl Theory of random matrices}, Les Houches Session LXI, 
eds.  E. Akkermans, G. Montambaux, J.L. Pichard, J. Zinn-Justin 
(North-Holland Pub. Co., Amsterdam, 1994). 

\bibitem{c<1} 
M.~R.~Douglas and S.~H.~Shenker,
Nucl.\ Phys.\ B {\bf 335}, 635 (1990);\\
D.~J.~Gross and A.~A.~Migdal,
Phys.\ Rev.\ Lett.\  {\bf 64}, 127 (1990);\\
D.~J.~Gross and A.~A.~Migdal,
Phys.\ Rev.\ Lett.\  {\bf 64}, 717 (1990);\\
M.~R.~Douglas,
Phys.\ Lett.\ B {\bf 238}, 176 (1990).

\bibitem{c=1}  
D.~J.~Gross and N.~Miljkovic,
Phys.\ Lett.\ B {\bf 238}, 217 (1990);\\
E.~Brezin, V.~A.~Kazakov and A.~B.~Zamolodchikov,
Nucl.\ Phys.\ B {\bf 338}, 673 (1990);\\
P.~Ginsparg and J.~Zinn-Justin,
Phys.\ Lett.\ B {\bf 240}, 333 (1990);\\
G.~Parisi,
Phys.\ Lett.\ B {\bf 238}, 209 (1990).

\bibitem{ikkt}
N.~Ishibashi, H.~Kawai, Y.~Kitazawa and A.~Tsuchiya,
Nucl.\ Phys.\ B {\bf 498}, 467 (1997)
[hep-th/9612115].

\bibitem{bfss}
T.~Banks, W.~Fischler, S.~H.~Shenker and L.~Susskind,
Phys.\ Rev.\ D {\bf 55}, 5112 (1997)
[hep-th/9610043].

\bibitem{uvir} 
S.~Minwalla, M.~Van Raamsdonk and N.~Seiberg,
JHEP {\bf 0002}, 020 (2000)
[hep-th/9912072].

\bibitem{Rey-talk}
S.-J.~Rey, talk given at "Strings 2001"-International Conference at 
Tata Institute for Fundamental Research (Mumbai, India) 
{\tt http://theory.theory.tifr.res.in/strings/Proceedings/\#rey-t}.

\bibitem{sentachyon}
A.~Sen,
JHEP {\bf 9808}, 012 (1998)
[hep-th/9805170];\\
A.~Sen,
JHEP {\bf 9912}, 027 (1999)
[hep-th/9911116];\\
A.~Sen and B.~Zwiebach,
JHEP {\bf 0003}, 002 (2000)
[hep-th/9912249].

\bibitem{bmn}
G.~Bhanot, G.~Mandal and O.~Narayan,
Phys.\ Lett.\ B {\bf 251}, 388 (1990).

\bibitem{VanRaamsdonk:2001jd}
M.~Van Raamsdonk,
[hep-th/0110093].

\bibitem{gww}
D.~J.~Gross and E.~Witten,
Phys.\ Rev.\ D {\bf 21} (1980) 446;\\
S.~Wadia, EFI-79/44-Chicago preprint (KEK Scanned Version); \\
S.~R.~Wadia,
Phys.\ Lett.\ B {\bf 93}, 403 (1980).

\bibitem{feynman} R.P. Feynman, 
{\sl Statistical Mechanics} (Benjamin Pub. Co., Boston, 1972).

\bibitem{sakita} B. Sakita, {\sl Quantum Theory of Many-Variable
Systems and Fields} (World Scientific Pub. Co., Singapore, 1985).

\bibitem{creutz}
M.~Creutz,
Rev.\ Mod.\ Phys.\  {\bf 50}, 561 (1978).

\bibitem{collective} 
A.~Jevicki and B.~Sakita,
Nucl.\ Phys.\ B {\bf 165}, 511 (1980);\\
S.~R.~Das and A.~Jevicki,
Mod.\ Phys.\ Lett.\ A {\bf 5}, 1639 (1990);\\
A.~M.~Sengupta and S.~R.~Wadia,
Int.\ J.\ Mod.\ Phys.\ A {\bf 6}, 1961 (1991);\\
D.~J.~Gross and I.~R.~Klebanov,
Nucl.\ Phys.\ B {\bf 352}, 671 (1991).

\bibitem{grossklebanov} 
D.~J.~Gross and I.~Klebanov,
Nucl.\ Phys.\ B {\bf 344}, 475 (1990).

\bibitem{dmw}
A.~Dhar, G.~Mandal and S.~R.~Wadia,
Mod.\ Phys.\ Lett.\ A {\bf 7}, 3129 (1992)
[hep-th/9207011];\\
{\sl ibid}.
Mod.\ Phys.\ Lett.\ A {\bf 8}, 3557 (1993)
[hep-th/9309028].

\bibitem{Ambjorn:2001xs}
J.~Ambjorn, K.~N.~Anagnostopoulos, W.~Bietenholz, F.~Hofheinz and J.~Nishimura,
[hep-th/0104260].


\bibitem{braneantibrane} 
O.~Aharony and M.~Berkooz,
Nucl.\ Phys.\ B {\bf 491}, 184 (1997)
[hep-th/9611215];\\
G. Lifschytz and S. Mathur,
Nucl.\ Phys.\ B {\bf 507}, 621 (1997)
[hep-th/9612087].

\bibitem{ddbar}
G. Mandal and S.R. Wadia,
Nucl.\ Phys.\ B {\bf 599}, 137 (2001)
[hep-th/0011094];\\
P. Kraus, A. Rajaraman and S.H. Shenker, .
Nucl.\ Phys.\ B {\bf 598}, 169 (2001)
[hep-th/0010016].

\bibitem{reys} Y.~Kiem, S.-J.~Rey, H.-T.~Sato and J.-T.~Yee,
[hep-th/0106121];\\
Y.~Kiem, S.-J.~Rey, H.-T.~Sato and J.-T.~Yee,
[hep-th/0107106];\\
Y.~Kiem, S.-S.~Kim, S.-J.~Rey and H.-T.~Sato,
[hep-th/0110066];\\
Y.~Kiem, S.~Lee, S.-J.~Rey and H.-T.~Sato,
[hep-th/0110215].

\bibitem{owl}
N.~Ishibashi, S.~Iso, H.~Kawai and Y.~Kitazawa,
Nucl.\ Phys.\ B {\bf 573}, 573 (2000)
[hep-th/9910004];\\
S.-J.~Rey and R.~von Unge,
Phys.\ Lett.\ B {\bf 499}, 215 (2001)
[hep-th/0007089];\\
S.~R.~Das and S.-J.~Rey,
Nucl.\ Phys.\ B {\bf 590}, 453 (2000)
[hep-th/0008042];\\
D.~J.~Gross, A.~Hashimoto and N.~Itzhaki,
[hep-th/0008075];\\
A.~Dhar and S.~R.~Wadia,
Phys.\ Lett.\ B {\bf 495}, 413 (2000)
[hep-th/0008144].

\bibitem{rey}
S.-J. Rey, {\sl Exact Answers to Approximate Questions: Noncommutative 
Dipole, Open Wilson Line, and UV-IR Duality}, Proceedings of `New Ideas in
String Theory', APCTP-KIAS Workshop (June, 2001) and 
of `Gravity, Gauge Theories, and Strings', Les Houches Summer School 
(August, 2001), to appear. 

\end{thebibliography}
\end{document}